\documentclass[amsmath,amssymb,aps,prx,twocolumn,nofootinbib,notitlepage,superscriptaddress,longbibliography]{revtex4-2}%
\usepackage{hyperref}
\usepackage{graphicx}
\usepackage{color}
\usepackage{comment}
\usepackage{bm}

\newcommand{\beq}{\begin{equation}}
\newcommand{\eeq}{\end{equation}\par\noindent}

\DeclareMathOperator{\vect}{vec}
\DeclareMathOperator{\rank}{rank}
\DeclareMathOperator{\tr}{Tr}

\begin{document}

\title{Experimentally bounding deviations from quantum theory in the landscape of generalized probabilistic theories}

\author{Michael D. Mazurek}
\affiliation{Institute for Quantum Computing and Department of Physics \& Astronomy, University of Waterloo, Waterloo, Ontario N2L 3G1, Canada}
\author{Matthew F. Pusey}
\altaffiliation{Present address: Department of Mathematics, University of York, Heslington, York YO10 5DD, UK}
\affiliation{Perimeter Institute for Theoretical Physics, 31 Caroline Street North, Waterloo, Ontario N2L 2Y5, Canada}
\affiliation{Department of Computer Science, University of Oxford, Wolfson
Building, Parks Road, Oxford OX1 3QD, UK}
\author{Kevin J. Resch}
\affiliation{Institute for Quantum Computing and Department of Physics \& Astronomy, University of Waterloo, Waterloo, Ontario N2L 3G1, Canada}
\author{Robert W. Spekkens}
\affiliation{Perimeter Institute for Theoretical Physics, 31 Caroline Street North, Waterloo, Ontario N2L 2Y5, Canada}

\begin{abstract} 
Many experiments in the field of quantum foundations seek to adjudicate between quantum theory and speculative alternatives to it.
This requires one to analyze the experimental data in a manner that does not presume the correctness of the quantum formalism. 
The mathematical framework of generalized probabilistic theories (GPTs)
provides a means of doing so.  We  present  a scheme for determining which GPTs are consistent with a given set of  experimental data.  It proceeds by performing tomography on the preparations and measurements in a self-consistent manner, i.e., without presuming a prior characterization of either. 
We illustrate the scheme by analyzing experimental data for a large set of preparations and measurements on the polarization degree of freedom of a single photon. 
We first test various hypotheses for the dimension of the GPT vector space for this degree of freedom. Our analysis identifies the most plausible hypothesis to be dimension 4, which is the value predicted by quantum theory. 
Under this hypothesis, we can draw the following additional conclusions from our scheme: 
(i) that the smallest and largest GPT state spaces that could describe photon polarization
 are a pair of polytopes, each approximating the shape of the Bloch Sphere and
having a volume ratio of $0.977 \pm 0.001$, which provides a quantitative bound on the scope for deviations from the state and effect spaces predicted by quantum theory, and (ii) that
 the maximal violation of the CHSH inequality can be at most $1.3\% \pm 0.1$ greater than the maximum violation allowed by quantum theory,
 and the maximal violation of a particular inequality for universal noncontextuality can not differ from the quantum prediction by more than this factor on either side. 
The only possibility for a {\em greater} deviation from the quantum state and effect spaces or for {\em greater} degrees of supra-quantum nonlocality or contextuality, according to our analysis, is if a future experiment (perhaps following the scheme developed here) discovers that additional dimensions of GPT vector space are required  to describe photon polarization, in excess of the 4 dimensions predicted by quantum theory to be adequate to the task. 
\end{abstract}

\maketitle
\tableofcontents

\section{Introduction}

Despite the empirical successes of quantum theory, it may one day be supplanted by a novel, post-quantum theory.\footnote{The fact that it has not yet been unified with general relativity, for instance, is often cited as evidence for this claim.
}  
Many researchers have sought to anticipate what such a theory might look like based on theoretical considerations, in particular,
by exploring how various natural physical principles narrow down the 
scope of possibilities
 in the landscape of all physical theories 
 (see~\cite{Chiribella2016} and references therein). In this article, we consider a complementary problem: how to narrow down the scope of possibilities directly from experimental data.

Most experiments in the field of quantum foundations aim
to adjudicate between quantum theory and some speculative alternative to it. 
They seek to constrain (and perhaps uncover) deviations from the quantum predictions. Although a few proposed alternatives to quantum theory can be articulated within the quantum formalism itself, such as models which posit intrinsic decoherence~\cite{ghirardi86,Percival89,milburn91,adler2000}, most are more radical. Examples include Almost Quantum Theory~\cite{navascues15,sainz17}, theories with higher-order interference~\cite{sorkin94,sinha10,hickman11,sollner12,park12,kauten17} (or of higher-order in the sense of Ref.~\cite{hardy01}), and modifications to quantum theory involving the quaternions~\cite{peres79,adler94,adler95,barnum2016composites}.

In order to assess whether experimental data provides any evidence for a given proposal (and against quantum theory), it is clearly critical that one {\em not} presume the correctness of quantum theory in the analysis.  Therefore it is inappropriate to use the quantum formalism to model the experiment.  A more general formalism is required.
Furthermore, it would be useful if rather than implementing dedicated experiments for each proposed alternative to quantum theory, one had a technique for 
directly determining the experimentally viable regions in the landscape of all possible physical theories.
The framework of generalized probabilistic theories (GPTs) provides the means to meet both of these challenges.

This framework adopts an operational approach to describing the content of a physical theory.  It has been developed over the past fifteen years in the field of quantum foundations (see \cite{hardy01,barrett07,chiribella10,chiribella11}, as well as  \cite{bengtsson06,hardyfoliable09,dakic09,dariano10,boxworld,masanes11,janotta14,barnum16,sainz17}),
continuing a long tradition of such approaches~\cite{mackey1963mathematical,ludwig1954grundlagen,ludwig1983problem, kraus1983states}.
It is {\em operational} because it takes the content of a physical theory to be merely what it predicts for the 
probabilities of outcomes of measurements in an experiment.

The GPT framework makes only very weak assumptions which are arguably unavoidable if an operationalist's conception of an experiment is to be meaningful.
 One is that experiments have a modular form, such that one part of an experiment can be varied independently of another, such as preparations and measurements for instance; another is that it is possible to repeat a given experimental configuration in such a way that it constitutes an i.i.d. source of statistical data.  
Beyond this, however, it is completely general.  
It has been used extensively to provide a common language for describing and comparing abstract quantum theory, classical probability theory, and many foils to these, including quantum theory over the real or quaternionic fields~\cite{barnum2016composites}, theories with higher-order interference~\cite{barnum2014higher,dakic14,Lee2017}, and the generalized no-signalling theory (also known as Boxworld)~\cite{barrett07,boxworld}.

 Using this framework, we propose a technique for analyzing experimental data that
allows researchers to overcome their implicit quantum bias---the tendency of  viewing all experiments through the lens of quantum concepts and the quantum formalism---and take a theory-neutral perspective on the data. 

Despite the fact that the GPT formalism is ideally suited to the task, to our knowledge,  it has not previously been applied to the analysis of experimental data (with the exception of Ref.~\cite{mazurek2016experimental}, which applied it to an experimental test of universal noncontextuality and which inspired the present work).

In this paper, we aim to answer the question: given specific experimental data, how does one find the set of GPTs that could have generated the data?  We call this the ``GPT inference problem''. 
Solving the problem requires implementing the GPT analogue of quantum tomography.  
Quantum tomography experiments  that have sought to characterize unknown states have typically presumed that the measurements are already well-characterized~\cite{vogel89,smithey93,haffner05,leibfried05,james01,dunn95,lvovsky09}, and those that have sought to characterize unknown measurements have typically presumed that the states are known~\cite{fiurasek01,lundeen09}. 
If one has no prior knowledge of either the states or the measurements, then one requires a tomography scheme that can characterize them both based on their interplay.  We call such a tomographic scheme {\em self-consistent}.  To solve the GPT inference problem, we introduce such
a self-consistent tomography scheme within the framework of GPTs.

We also illustrate the use of our technique with an experiment on the polarization degree of freedom of a single photon.  For each of a large number of preparations, we perform a large number of measurements, and we analyze the data using 
 our self-consistent tomography scheme to infer a GPT characterization of both the preparations and the measurements.

To clarify what, precisely, our analysis implies, we begin by distinguishing two ways in which nature might deviate from the predictions of quantum theory within the framework of GPTs.
The first possibility is that it exhibits a deviation (relative to what quantum theory predicts for the system of interest) in the particular shapes of the spaces of GPT state vectors and GPT effect vectors but {\em no deviation} in the dimensionality of the GPT vector space. 
The second possibility is that it deviates from quantum expectations even in the dimensionality. 
 
 From our experimental data, we find no evidence of either sort of deviation. If nature does exhibit deviations and these are of the first type (i.e., deviations to shapes but not to dimensions), then we are able to put quantitative bounds on the degree of such deviations.
 If nature exhibits deviations of the second type  
  (dimensional deviations), then although our GPT inference technique may fail to detect them in a given experiment, 
 it does provide an opportunity for doing so.  In the next few paragraphs, we try to explain the precise sense in which there is such an opportunity.
 
If dimensional deviations from quantum theory happen to only be
significant for some exotic new types of preparations and measurements, then insofar as our experiment only probes a photon's polarization in conventional ways (using waveplates and beamsplitters),
 there is nothing in its design ensuring that such deviations will be found.  
Nonetheless, it is still the case that our experiment (and any other that implements our technique on data obtained by probing a system in conventional ways) has an opportunity to discover such deviations, even in the absence of any knowledge of the type of exotic procedures required to make such deviations significant.
To see why this is the case, note that there are  two ways in which an experiment might discover new physics: the ``terra nova'' strategy, wherein one's experiment probes a new phenomenon or regime of some physical quantity, and the  ``precision'' strategy, wherein one's experiment  achieves increased precision for a previously explored phenomenon or regime.

To illustrate the distinction, consider a counterfactual history of physics, wherein the special theory of relativity 
 was not discovered by theoretical considerations
  but was instead inferred primarily from experimental discoveries.  Imagine, for instance, that 
it began with 
 the discovery of corrections to the established (nonrelativistic) formulas for properties of moving bodies, such as the expression for their kinetic energy or the Doppler shift of the radiation they emit. 
On the one hand, 
an experimenter who, for whatever reason, had found herself investigating the behaviour of systems accelerated to speeds 
that were a significant fraction of the speed of light (without necessarily even knowing that the speed of light was a limit) 
would have found 
significant deviations from various nonrelativistic formulas.  On the other hand, 
an experimenter who probed systems at unexceptional speeds, (i.e., speeds {\em small} compared to the speed of light), but with a degree of precision much higher than had been previously achieved could still have discovered the inadequacy of nonrelativistic formulas by detecting small but statistically significant deviations from these.

The experiment we report provides an opportunity to discover a 
deviation (from quantum theory) in the dimension of the GPT vector space required to describe photon polarization
 because it provides a precision characterization of a large set of preparations and measurements thereon. 
If experimental set-ups designed to realize conventional preparations and measurements inadvertently extend some small distance into the space of exotic preparations and measurements, say, by fluctuations or small systematic effects,  then our technique can reveal this fact by showing that the expected dimensionality for the GPT vector space does not fit the data.  The full scope of possible preparations and measurements for photon polarization might be radically different from what our quantum expectations dictate (incorporating new exotic procedures), and yet one could, by serendipity, experimentally realize a set of preparations and measurements that are tomographically complete for this full set rather than being merely sufficient for characterizing the conventional procedures.  In other words, the realized set could manage to span the full post-quantum GPT vector space in spite of their not having been designed to do so. 
In Section \ref{sc:tomocomplevidence}, we provide a more detailed discussion of this point.\footnote{In particular, we note there that an experiment which was designed to realize only ``rebit'' states and effects (in the Bloch representation, these have no component of the $Y$ Pauli operator) could easily end up inadvertently realizing nonrebit states and effects (e.g., by incorporating a small and inadvertent admixture of the $Y$ Pauli operator).}

Applying our GPT inference technique to our experimental data, we find 
that our experiment is best represented by a GPT of dimension 4, which is what quantum theory predicts to be the appropriate dimension for photon polarization. 
In other words, we find no evidence for a deviation in the dimension of the GPT vector space, relative to quantum expectations, at the precision frontier using conventional means of probing photon polarization.   We can therefore conclude that one of the following possibilities must hold: (i) there are no dimensional deviations, (ii) there are dimensional deviations, which exotic preparations and measurements would reveal, but the procedures realized in our experiment contain strictly no exotic component, (iii) there are dimensional deviations, which exotic preparations and measurements would reveal, and the procedures realized in our experiment do contain some exotic component, but the latter is not visible at the level of precision achieved in our experiment.

We now describe what further conclusions we can draw from our experiment supposing that the realized preparations and measurements in our experiment are {\em tomographically complete}, that is, supposing that they have nontrivial components in all dimensions of the GPT vector space describing photon polarization and that these components are visible at the level of precision achieved in our experiment.  In other words, we now describe what further conclusions we can draw from our experiment if we suppose that it is possibility (i), rather than possibilities (ii) or (iii), that holds.
 In this case, we are able to place bounds (at the 1\% level) on how much the 
state and effect spaces of the true GPT might deviate from those predicted by quantum theory. 
   In addition, we are able to draw explicit quantitative conclusions about three types of such putative deviations, which we now outline.

The {\em no-restriction hypothesis}~\cite{chiribella10} asserts that if some measurement is logically possible (i.e., it gives positive probabilities for all states in the theory) then it should be physically realizable. It is true of quantum theory---indeed, it is a popular axiom in many axiomatic reconstructions thereof. A failure of the no restriction hypothesis, therefore, constitutes a departure from quantum theory.  We put quantitative bounds on the possible degree of this failure, that is, on the potential gap between the set of measurements that are physically realizable and those that are logically possible.  Recalling the scope of possible conclusions (i)-(iii) above, the only way for any future experiment to overturn this conclusion about deviations from the no-restriction hypothesis is if it demonstrated the need for dimensional deviations. 

We can also put an upper bound on the amount by which nature might violate Bell inequalities in excess of the amount predicted by quantum theory.  Specifically, for the CHSH inequality~\cite{clauser69}, we show that, for photon polarization, any greater-than-quantum degree of violation is no more than $1.3\%\pm 0.1$ higher than the quantum bound.  To our knowledge, this is the first proposal for how to obtain an experimental {\em upper} bound on the degree of Bell inequality violation in nature.     The only possibility for a future experiment on photon polarization to violate the quantum bound by more than $1.3\%\pm 0.1$ is if it demonstrated the need for dimensional deviations. \color{black}

In a similar vein, we consider noncontextuality inequalities.  These are akin to Bell inequalities, but test the hypothesis of universal noncontextuality~\cite{spekkens05} rather than local causality.  Here, our technique provides both an upper and a lower bound on the degree of violation.  For a particular noncontextuality inequality, described in Ref.~\cite{spekkens09},  we find that the true value of the violation is no more than $1.3\%\pm 0.1$ higher and no less than $1.3\%\pm 0.1$ lower than the quantum bound.    As with Bell inequalities, the only way for any future experiment on photon polarization to find a violation outside this range is if it demonstrated the need for dimensional deviations. \color{black}

Although we have {\em not} here sought to  implement any terra nova strategy for finding deviations from quantum theory, any future experiment that aims to do so can make use of our GPT inference technique to analyze the data and evaluate the evidence.
Inasmuch as terra nova strategies, relative to precision strategies, provide a complementary (and presumably better) opportunity for finding new physics, our GPT  inference technique is also significant insofar as it provides the means to analyze such experiments.

\section{The framework of generalized probabilistic theories}

\subsection{Basics}\label{basics}

For any system, in any physical theory, there will in general be many possible ways for it to be prepared, transformed, and measured.  Here, each preparation procedure, transformation procedure and measurement procedure is conceived as a list of instructions for what to do in the laboratory.  The different combinations of possibilities for each procedure defines a collection of possible experimental configurations.  We will here restrict our attention to experimental configurations of the prepare-and-measure variety: these are the configurations where there is no transformation intervening between the preparation and the measurement and where the measurement is terminal (which is to say that the system does not persist after the measurement).
We further restrict our attention to binary-outcome measurements.

A GPT aims to describe only the operational phenomenology of a given experiment.  In the case of a prepare-and-measure experiment, it aims to describe only the relative probabilities of the different outcomes of each possible measurement procedure when it is implemented following each possible preparation procedure.   For binary-outcome measurements, it suffices to specify the probability of one of the outcomes since the other is determined by normalization.  If we denote the outcome set $\{0,1\}$, then it suffices to specify the probability of the event of obtaining outcome 0 in measurement $M$.  This event will be termed an {\em effect} and denoted $[0|M]$.  
 
Thus a GPT specifies a probability $p(0|P,M)$ for each preparation $P$ and measurement $M$.  
Denoting the cardinality of the set of all preparations (respectively all measurements) by $m$ (respectively $n$), the set of these probabilities can be organized into an $m\times n$ matrix, denoted $D$, where the rows correspond to distinct preparations and the columns correspond to distinct effects, 
\[
D \equiv \left(
\begin{array}{cccc}
p(0|P_1,M_1)  & p(0|P_1,M_2)  & \cdots & p(0|P_1,M_n)    \\
p(0|P_2,M_1)  & p(0|P_2,M_2)  & \cdots  & p(0|P_2,M_n) \\
\cdots & \cdots  & \cdots  \\
p(0|P_m,M_1)  & p(0|P_m,M_2)  & \cdots   & p(0|P_m,M_n)
\end{array}
\right)
\]
 We refer to $D$ as the 
  {\em probability matrix}  associated to the physical theory.  Because it specifies the probabilities for {\em all} possibilities for the preparations and the measurements, it contains all of the information about the putative physical theory for prepare-and-measure experiments.\footnote{Note that although the presentation as a table suggests that the sets of preparations and measurements are discrete, there could in fact be a continuum of possibilities for each set.  If the continuous variable labelling the preparations in the theory is $x$ and that labelling the measurements in the theory is $y$, then the complete information about the physical theory is given by the function $f(x,y) := p(0|P_x,M_y)$.   The GPT is a theoretical abstraction, so it is acceptable if it is presumed to contain such continua.} 

Defining
\[
k \equiv {\rm rank}(D) 
\]
then one can factor $D$ into a product of two rectangular matrices, 
\begin{equation} 
 D = S E
 \label{decomp}
 \end{equation}
where $S$ is an $(m \times k)$ matrix and $E$ is a $(k \times n)$ matrix. 

Denoting the $i$th row of $S$ by the row vector ${\bf s}^T_{P_i}$ (where $T$ denotes transpose) and the $j$th column of $E$ by the column vector ${\bf e}_{[0|M_j]}$, we can write
\begin{align}
D &= 
\left(
\begin{array}{c}
{\bf s}^T_{P_1}   \\
{\bf s}^T_{P_2} \\
\cdots   \\
{\bf s}^T_{P_m}
\end{array}
\right)
\left(
\begin{array}{cccc}
{\bf e}_{[0|M_1]} & {\bf e}_{[0|M_2]}  & \cdots & {\bf e}_{[0|M_n]}   
\end{array}
\right),
\end{align}
so that 
\begin{equation}
p(0|P_i,M_j) = {\bf s}_{P_i} \cdot {\bf e}_{[0|M_j]}.
\end{equation}
Factoring $D$ in this way allows us to associate to each preparation $P$ a $k$-dimensional vector ${\bf s}_{P}$ and to each effect $[0|M]$ a $k$-dimensional vector ${\bf e}_{[0|M]}$ such that the probability of obtaining the effect $[0|M]$ on the preparation $P$ is recovered as their inner product, $p(0|P,M) = {\bf s}_{P} \cdot {\bf e}_{[0|M]}$.  The vectors ${\bf s}_{P}$ and ${\bf e}_{[0|M]}$ will be termed {\em GPT state vectors} and {\em GPT effect vectors} respectively.  A particular GPT is specified by the sets of all allowed GPT state and effect vectors, denoted by $\mathcal{S}$ and $\mathcal{E}$, respectively.

Because the $n$ GPT effect vectors associated to the set of all measurement effects lie in a $k$-dimensional vector space, only $k$ of them are linearly independent. Any set of $k$ measurement effects whose associated GPT effect vectors form a basis for the space will be termed a {\em tomographically complete} set of measurement effects. The terminology stems from the fact that if one seeks to deduce the GPT state vector of an unknown preparation from the probabilities it assigns to a set of characterized measurement effects (the GPT analogue of quantum state tomography) then  this set of GPT effect vectors must form a basis of the $k$-dimensional space. 
Similarly, any set of $k$ preparations whose associated GPT state vectors form a basis for the space will be termed {\em tomographically complete} because  to deduce the GPT effect vector of an unknown measurement effect from the probabilities assigned to it by a set of known preparations, the GPT state vectors associated to the latter must form a basis.

For any GPT, we necessarily have that the rank of $D$ satisfies $k \le \min\{m,n\}$, but in general, we expect $k$ to be much smaller than $m$ or $n$. 

There is a freedom in the decomposition of Eq.~\eqref{decomp}.  Specifically, for any invertible $(k\times k)$ matrix $R$, we have $D = SE = (SR^{-1})(RE)$.  Thus, there are many decompositions of $D$ of the type described. The vectors $\{ {\bf s}_{P_i}\}_i$ and $\{ {\bf e}_{[0|M_j]} \}_j$  depend on the specific decomposition chosen. However, for any two choices of decompositions $SE$ and $S'E'$, the vectors $\{ {\bf s}_{P_i}\}_i$ and $\{ {\bf s}'_{P_i}\}_i$  (and the vectors  $\{ {\bf e}_{[0|M_j]} \}_j$ and  $\{ {\bf e}'_{[0|M_j]} \}_j$) are always related by a linear transformation.

Note that any basis of the $k$-dimensional vector space remains so under a linear transformation, so the property of being tomographically complete is independent of the choice of representation.

It is worth noting that for {\em any} physical theory, the GPT framework provides a complete description of its operational predictions for prepare-and-measure experiments.  In this sense, the GPT framework is completely general.  Furthermore, one can show that under a very weak assumption it provides the most efficient description of the theory, in the sense that it is a description with the smallest number of parameters.  The weak assumption is that it is possible to implement arbitrary convex mixtures of preparations without altering the functioning of each preparation in the mixture, so that for any set of GPT state vectors that are admitted in the theory, all of the vectors in their convex hull are also admitted in the theory.   See Theorem 1 of Ref.~\cite{hardyfoliable09} for the proof.

 We will here make this weak assumption and restrict our attention to GPTs wherein  any convex mixture of  preparation procedures is another valid preparation procedure, so that the set of GPT state vectors is convex~\cite{hardy01}. 
We refer to the set $\mathcal{S}$ of GPT states in a theory as its {\em GPT state space}.
We also make the weak assumption that any convex mixture of measurements and any classical post-processing of a measurement is another valid measurement.  This implies that the set of GPT effect vectors consists of the intersection of two cones, which can be described as follows: there is some set of ray-extremal GPT effect vectors, such that the first cone is the convex hull of all positive multiples of these vectors, and the second cone is the set of vectors which can be summed with a vector in the first cone to yield the unit effect vector $\mathbf{u}$ (defined below).  (This ensures that if a given effect $\mathbf{e}$ is in the GPT, then so is the complementary effect $\bar{\mathbf{e}} :=\mathbf{u} - \mathbf{e}$.) We will use the term ``diamond'' to describe this sort of intersection of two cones, and we refer to the set $\mathcal{E}$ of GPT effects in a theory as its {\em GPT effect space}. 

It is worth noting that although GPTs which fail to be closed under convex mixtures and classical post-processing are of theoretical interest --- there are interesting foils to quantum theory of this type~\cite{spekkens05,Schumacher2016} --- one does not expect them to be candidates for the true GPT describing nature because there seems to be no obstacle in practice to mixing or post-processing procedures in an arbitrary way.  To put it another way, the evidence suggests that the GPT describing nature must include classical probability theory as a subtheory, thereby providing the resources for implementing arbitrary mixtures and post-processings.

Distinct physical theories (i.e., distinct GPTs) are distinguished by the {\em shapes} of the GPT state space and the GPT effect space, where these shapes are defined up to a linear transformation, as described earlier. 

We end by highlighting some conventions we adopt in representing GPTs.  Define the {\em unit measurement effect} as the one which occurs with probability 1 for all preparations (it is represented by a column of 1s in $D$), and denote it by $\mathbf{u}$. 
 Because each $\mathbf{s}_{P}$ will have an inner product of 1 with $\mathbf{u}$ (by normalization of probability), it follows that there are only $k-1$ free parameters in the GPT state vector.  
We make a conventional choice (i.e., a particular choice within the freedom of linear transformations)
 to represent the unit effect by the GPT effect vector $(1,0,0,\dots)^T$.  This choice forces the first component of all of the GPT state vectors to be 1.  In this case, one can restrict the search for factorizations $D = SE$ to those for which the first column of $S$ is a column of $1$s.  It also follows that 
 the projection of all  GPT state vectors along one of the axes of the $k$-dimensional vector space has value 1, and consequently it is useful to only depict the projection of the GPT state vectors into the complementary $(k{-}1)$-dimensional subspace.

 \subsection{Examples}\label{examples}
 
Some simple examples serve to clarify the notion of a GPT. First, consider a 2-level quantum system (qubit).  The set of all preparations is represented by the set of all positive trace-one operators on a 2-dimensional complex Hilbert space, that is, $\rho \in \mathcal{L}(\mathbb{C}^2)$ with $\mathcal{L}$ denoting the linear operators, such that $\rho \ge 0$ and $\tr(\rho) =1$.
Each measurement effect is associated with a positive operator less than identity, 
$0 \le Q \le \mathbb{I}$.
Each measurement effect and each preparation can also be represented by a vector in a real 4-dimensional vector space by simply decomposing the operators representing them relative to any orthonormal basis of Hermitian operators.  The Born rule is reproduced by the vector space inner product because it is simply the inner product of the associated operators relative to the Hilbert-Schmidt norm. 

\begin{figure}
  \centering
  \includegraphics[]{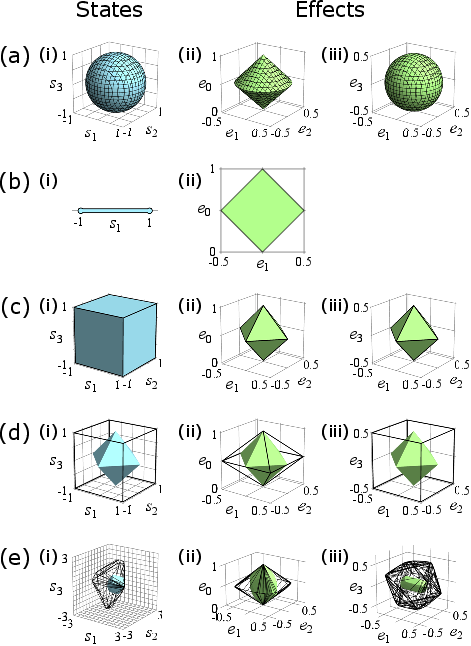}
  \caption{(Color.) Some paradigm examples of GPTs. The solid shapes represent the true state and effect spaces for that GPT, while the black wireframe shapes represent the duals of these (for the duality relation described in Sec.~\ref{sc:dualspaces}).
  (i) the true state space (solid blue) and the space of 
logically possible states (wireframe).
   (ii)-(iii) The true effect space (solid green) and the space of logically possible effects (wireframe). 
   For the cases where $k=4$, the effect spaces are 4d, and we depict them by a pair of 3d projections.
     (a) A qubit ($k=4)$.  (b) A classical bit $(k=2)$.  (c) The $k=4$ system in Boxworld. {(d)} The convex closure of the Spekkens toy theory for the simplest system $(k=4)$. (e) A generic GPT with $k=4$, obtained from a randomly generated rank-4 matrix of probabilities.}
  \label{fg:gptspaces}
\end{figure}

The most common example of such a representation is the one that uses (a scalar multiple of) the four Pauli operators, $\{ \frac{1}{2} \mathbb{I}, \frac{1}{2} \sigma_x, \frac{1}{2} \sigma_y, \frac{1}{2} \sigma_z\}$, as the orthonormal basis of the space of operators.   
A preparation represented by a density operator $\rho$ is associated with the 4-dimensional real vector $\mathbf{s}\equiv (s_0,s_1,s_2,s_3)$, via the relation $\rho = \frac{1}{2} \mathbf{s} \cdot \bm{\sigma}$, where $\bm{\sigma}\equiv (\mathbb{I}, \sigma_x, \sigma_y, \sigma_z)$, or equivalently, $\rho = \frac{1}{2}\left( s_0 \mathbb{I} + s_1 \sigma_x + s_2 \sigma_y + s_3 \sigma_z \right)$.  The condition $\tr(\rho)=1$ implies that $s_0=1$, and the conditions $\tr(\rho)=1$ and $\rho \geq 0$ together imply that $\sqrt{s_1^2+s_2^2+s_3^2} \leq 1$. Consequently, there is only a 3-dimensional freedom in specifying a quantum state.  Geometrically, the possible $\mathbf{s}$ describe a ball of radius 1, conventionally termed the Bloch Sphere\footnote{Strictly speaking, however, it should be called the Bloch Ball.} and depicted in Fig.~\ref{fg:gptspaces}(a)(i).
A measurement effect represented by an operator $Q$ 
 is associated with the 4-dimensional real vector $\mathbf{e}\equiv (e_0 ,e_1,e_2,e_3)$, via the relation $Q = \mathbf{e} \cdot \bm{\sigma}$.
The conditions $Q\ge 0$ and $Q \le \mathbb{I}$ imply that $0\leq e_0\leq 1$,  $\sqrt{e_1^2+e_2^2+e_3^2} \leq e_0$ and $\sqrt{e_1^2+e_2^2+e_3^2} \leq 1- e_0$, which
constrains $\mathbf{e}$ to lie within the intersection of two four-dimensional cones, which we refer to as the Bloch Diamond and depict via a pair of three-dimensional projections in Fig.~\ref{fg:gptspaces}{(a)}(ii)-(iii)\footnote{\label{blochconvention}Note that the relation we assume to hold between a qubit measurement effect $Q$ and the Bloch vector $\mathbf{e}$ representing it, namely, $Q=\mathbf{e}\cdot\bm{\sigma}$, differs from the standard convention used in quantum information theory by a factor of $\frac{1}{2}$. Our choice of convention ensures the GPT effect vectors are equal to the Bloch vectors, whereas in the standard convention there would be a factor of $\frac{1}{2}$ difference between the two.
}. 

As noted in the discussion of the GPT framework, this geometric representation of the quantum state and effect spaces is only one possibility among many. 
 If we define a linear transformation of the state space by any invertible $4\times4$  matrix and we take the corresponding inverse linear transformation on the effect space, the new state and effect spaces will also provide an adequate representation of all prepare-and-measure experiments on a single qubit.  (Note that implementing a linear transformation of this form is equivalent to representing quantum states and effects with respect to a different basis of Hermitian operators.)

Classical probabilistic theories can also be formulated within the GPT framework.   Consider the simplest case of a classical system with two possible physical states, i.e., a classical bit,  for which  $k=2$.
 The set of possible preparations of this system is simply the set of normalized probability distributions on a bit , $\vec{\mu} = (\mu_0,\mu_1)$, where $0\le \mu_0,\mu_1 \le 1$ and $\mu_0 + \mu_1=1$.
The most general measurement effect is a pair of probabilities, specifying the probability of that effect occuring for each value of the bit, that is, $\vec{\xi} = (\xi_0,\xi_1)$ where $0\le \xi_0,\xi_1 \le 1$.  The probability of a particular measurement effect occuring when implemented on a particular preparation is clearly just the inner product of these, $\vec{\mu} \cdot \vec{\xi}$.
The positivity and normalization constraints imply that the convex set of state vectors describes a line segment from $(1,0)$ to $(0,1)$, and the set of effect vectors is the square region with vertices $(0,0), (1,0), (0,1)$ and $(1,1)$. 

For ease of comparison with our examples of GPTs, it is useful to consider a linear transformation of this representation, corresponding geometrically to a rotation by 45 degrees.  We represent each preparation by a state vector $\mathbf{s} = (1,s_1)$, where $-1\leq s_1 \leq 1$, and each measurement effect by an effect vector $\mathbf{e} = (e_0,e_1)$ where $-1/2 \le e_1 \le 1/2$ and $e_0 \geq |e_1|$ and $e_0 \leq 1 - |e_1|$ (with the experimental probabilities still given by their inner product, $\mathbf{s} \cdot \mathbf{e}$).  The convex set of these state vectors can then be depicted as a horizontal line segment, and the set of effect vectors by a diamond with a line segment at its base,  as in Fig.~\ref{fg:gptspaces}(b).  This representation makes it clear that the state and effect spaces of a classical bit are contained within those of a qubit (as the quantum states and effects whose representation as operators are diagonal in some fixed basis of the Hilbert space). 

One can also consider GPTs that are neither classical nor quantum. In the GPT known as ``Boxworld''~\cite{barrett07,boxworld} (originally called ``generalized no-signalling theory''), correlations can be stronger than in quantum theory, violating Bell inequalities by an amount in excess of the maximum quantum violation.   The $k=3$ system in Boxworld, known as the ``generalized no-signalling bit'', has received a great deal of attention.  A pair of such systems can generate the stronger-than-quantum correlations known as a Popescu-Rohrlich box~\cite{popescu1994quantum} from which the name Boxworld derives.   These achieve a CHSH inequality violation equal to the algebraic maximum.  Such correlations are achievable in Boxworld because there are some states that respond deterministically to multiple effects, and there are also some effects that respond deterministically to multiple states.  Boxworld also has a $k=4$ system, which shares features of the generalized no-signalling bit and is, in certain respects, more straightforward to compare to a qubit.  It is the latter that we depict in Fig.~\ref{fg:gptspaces}(c).

Another alternative to classical and quantum theories is the toy theory introduced by one of the authors~\cite{spekkens2007evidence}.  We here consider a variant of this theory, wherein one closes under convex combinations.  The simplest system has $k=4$ and has the state and effect spaces depicted in 
Figure~\ref{fg:gptspaces}(d).\footnote{These state and effect spaces are strictly contained within those of the classical theory for a system with four physical states (the $k=4$ system in the classical theory), which corresponds to the fact that the theory can be understood as the result of imposing an additional restriction relative to what can be achieved classically.}

Finally, Fig.~\ref{fg:gptspaces}(e) illustrates a generic example of a GPT with $k=4$. We constructed this GPT by generating a rank 4 matrix of random probabilities, and found GPT representations of the state and effect spaces from that.

In this paper, we describe a technique for estimating the GPT state and effect spaces that govern nature directly from experimental data.   The examples described above illustrate the diversity of forms that the output of our technique could take.  

\subsection{Dual spaces}\label{sc:dualspaces}

Finally, we review the notion of the dual spaces of GPT state  and effect spaces. We will call a vector $\mathbf{s}\in \mathbb{R}^k$ a \textit{logically possible state} if it assigns a valid probability to every measurement effect allowed by the GPT.  Mathematically, the space of logically possible states, denoted $\mathcal{S}_{\rm logical}$, contains all $\mathbf{s}\in \mathbb{R}^k$ such that  $\forall \mathbf{e} \in \mathcal{E}: 0 \leq \mathbf{s} \cdot  \mathbf{e}
 \ \leq 1$ and such that $ \mathbf{s} \cdot  \mathbf{u} \ = 1$.
  From this definition, it is clear that $\mathcal{S}_{\rm logical}$ is the intersection of the geometric dual of $\mathcal{E}$ and the hyperplane defined by $ \mathbf{s} \cdot  \mathbf{u} \ = 1$; as a shorthand, we will refer to $\mathcal{S}_{\rm logical}$ simply as ``the dual of $\mathcal{E}$'', and denote the relation by $\mathcal{S}_{\rm logical} \equiv {\rm dual}(\mathcal{E})$. Analogously, the set of logically possible effects, denoted $\mathcal{E}_{\rm logical}$, contains all $\mathbf{e}\in \mathbb{R}^k$ such that $\forall \mathbf{s} \in \mathcal{S}:  0 \leq \mathbf{s} \cdot {\mathbf e} \leq 1$.  Defining the set of subnormalized states by $\mathcal{\hat{S}} \equiv \{ w{\bf s} : {\bf s} \in \mathcal{S}, w\in [0,1]\}$,  $\mathcal{E}_{\rm logical}$ is the geometric dual of $\mathcal{\hat{S}}$.  For simplicity, we will refer to $\mathcal{E}_{\rm logical}$ simply as ``the dual of $\mathcal{S}$'', and denote the relation by  $\mathcal{E}_{\rm logical} \equiv {\rm dual}(\mathcal{S})$. 
 
 GPTs in which $\mathcal{S}_{\rm logical} = \mathcal{S}$ and $\mathcal{E}_{\rm logical} = \mathcal{E}$ (the two conditions are equivalent) are said to satisfy the {\em no-restriction hypothesis}~\cite{chiribella10}. In a theory that satisfies the no-restriction hypothesis, every logically allowed GPT effect vector corresponds to a physically allowed measurement, and (equivalently) every logically allowed GPT state vector corresponds to a physically allowed preparation.
  In theories wherein $\mathcal{S}_{\rm logical} \ne \mathcal{S}$ and $\mathcal{E}_{\rm logical} \ne \mathcal{E}$, by contrast, there are vectors that do not correspond to physically allowed states but nonetheless assign valid probabilities to all physically allowed effects, and there are vectors that do not correspond to physically allowed effects but are nonetheless assigned valid probabilities by all physically allowed states.  
 
For each of the examples in Fig.~\ref{fg:gptspaces}, we have depicted the dual to the effect space alongside the state space and the dual of the state space alongside the effect space, as wireframes.  Quantum theory, classical probability theory, and Boxworld provide examples of GPTs that satisfy the no-restriction hypothesis, as illustrated in Fig.~\ref{fg:gptspaces}(a),(b),(c), while the GPTs presented in Fig.~\ref{fg:gptspaces}(d),(e) are examples of GPTs that violate it.
 
\subsection{The GPT inference problem}\label{ModellingExptGPTFramework}
 
The true GPT state and effect spaces, $\mathcal{S}$ and $\mathcal{E}$, are theoretical abstractions, describing the full set of GPT state and effect vectors that could be realized in principle if one could eliminate all noise.   However, the ideal of noiselessness is never achieved.  Therefore, the GPT state and effect vectors describing the preparation and measurement effects realized in any experiment are necessarily bounded away from the extremal elements of $\mathcal{S}$ and $\mathcal{E}$.  Geometrically, the realized GPT state and effect spaces will be contracted relative to their true counterparts.

There is another way in which the experiment necessarily differs from the theoretical abstraction: it may be impossible for the set of experimental configurations in a real experiment to probe {\em all} possible experimental configurations allowed by the GPT. For instance, for quantum theory there are an {\em infinite} number of convexly extremal preparations and measurements even for a single qubit, while a real experiment can only implement a finite number of each.  

Because we assume convex closure,  the realized GPT state and effect spaces will be polytopes.   If the experiment probes a sufficiently dense sample of the preparations and measurements allowed by the GPT, then the shapes of these polytopes ought to resemble the shapes of their true counterparts.
 
We term the convex hull of the GPT states that are actually realized in an experiment the {\em realized GPT state space}, and denote it by $\mathcal{S}_{\rm realized}$.  
Because every preparation is noisier than the ideal version thereof, this will necessarily be {\em strictly} contained within the true GPT state space $\mathcal{S}$.  
Similarly, we term the diamond defined by the 
GPT measurement effects that are actually realized in an experiment the {\em realized GPT effect space}, and denote it $\mathcal{E}_{\rm realized}$.  Again, we expect it to be strictly contained within $\mathcal{E}$.
By dualization, $\mathcal{S}_{\rm realized}$ defines the set of GPT effect vectors that are logically consistent with the realized preparations, which we denote by $\mathcal{E}_{\rm consistent}$, that is, $\mathcal{E}_{\rm consistent} \equiv {\rm dual}(\mathcal{S}_{\rm realized})$. Similarly, the set of GPT state vectors that are logically consistent with the realized measurement effects is $\mathcal{S}_{\rm consistent} \equiv {\rm dual}(\mathcal{E}_{\rm realized})$.

\begin{figure*}[htb]
  \centering
  \includegraphics[]{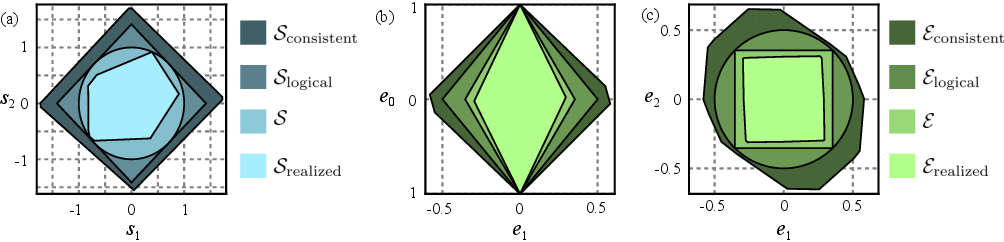}
  \caption{(Color). 
  An illustration of the inclusion relations among the different spaces of states and effects considered in this work.  We use a generic $k=3$ example for ease of depicting set inclusions. 
(a) The different spaces of states. (b),(c) The 2-d projections of the different spaces of effects. 
  The GPT specifies a space of true states, $\mathcal{S}$,  and effects, $\mathcal{E}$. From these, one can find the sets of logically possible states, $\mathcal{S}_{\rm logical}$, and effects $\mathcal{E}_{\rm logical}$. $\mathcal{E}_{\rm logical}$ is the dual of $\mathcal{S}$, and it represents all effects which return probabilities between 0 and 1 when applied to every possible state in $\mathcal{S}$. Similarly, $\mathcal{S}_{\rm logical}$ is the dual of $\mathcal{E}$. The logical state (effect) space must always contain the true state (effect) space. The spaces $\mathcal{S}_{\rm realized}$ and $\mathcal{E}_{\rm realized}$ are the GPT representations of the preparations and measurement effects actually realized in the experiment.  As any real experiment necessarily contains a finite amount of noise, $\mathcal{S}_{\rm realized}$ will always be contained within $\mathcal{S}$, and $\mathcal{E}_{\rm realized}$ will always be contained within $\mathcal{E}$.
 $\mathcal{E}_{\rm consistent}$ is the dual of $\mathcal{S}_{\rm realized}$ (and thus will always contain $\mathcal{E}_{\rm logical}$), and it represents all effects that are logically consistent with the set of states realized in the experiment. Similarly, $\mathcal{S}_{\rm consistent}$ will always contain $\mathcal{S}_{\rm logical}$ as it is the dual of $\mathcal{E}_{\rm realized}$.}
  \label{fg:fourspaces}
\end{figure*}

Suppose one has knowledge of the 
 realized GPT state and effect spaces $\mathcal{S}_{\rm realized}$ and $\mathcal{E}_{\rm realized}$ for some experiment. What can one then infer about $\mathcal{S}$ and $\mathcal{E}$?  The answer is that $\mathcal{S}$ can be any convex set of GPT states that lies strictly between $\mathcal{S}_{\rm realized}$ and $\mathcal{S}_{\rm consistent}$.  
For every such possibility for $\mathcal{S}$, $\mathcal{E}$ could be any diamond of GPT effects that lies between $\mathcal{E}_{\rm realized}$ and ${\rm dual}(\mathcal{S})  \subset \mathcal{E}_{\rm consistent}$. These inclusion relations are depicted in Fig.~\ref{fg:fourspaces}.

The larger the gap between $\mathcal{S}_{\rm realized}$ and $\mathcal{S}_{\rm consistent}$, the more choices of $\mathcal{S}$ and $\mathcal{E}$ there are that are consistent with the experimental data.  
An example helps illustrate the point.  Suppose that one found $\mathcal{S}_{\rm realized}$ and $\mathcal{E}_{\rm realized}$ to be the GPT state and effect spaces depicted in Fig.~\ref{fg:gptspaces}(d). In this case $\mathcal{S}_{\rm realized}$ is represented by the blue octahedron in Fig.~\ref{fg:gptspaces}(d)(i), and $\mathcal{E}_{\rm realized}$ is the green diamond with an octahedral base depicted in Fig.~\ref{fg:gptspaces}(d)(ii-iii). The wireframe cube in Fig.~\ref{fg:gptspaces}(d)(i) is the space of states $\mathcal{S}_{\rm consistent}$ that is the dual of $\mathcal{E}_{\rm realized}$, and the wireframe diamond with a cubic base in Fig.~\ref{fg:gptspaces}(d)(ii-iii) is the space of effects $\mathcal{E}_{\rm consistent}$ that is the dual of $\mathrm{S}_{\rm realized}$. 
Which GPTs are candidates for the true GPT in this case?  
The answer is: those whose state space contains the blue octahedron and is contained by the wireframe cube in Fig.~\ref{fg:gptspaces}(d)(i) and whose effect space contains the green diamond with the octohedral base in Fig.~\ref{fg:gptspaces}(d)(ii)-(iii) (the consistency of the effect space with the state space is a given if one grants that the pair is a valid GPT).  
By visual inspection of Fig.~\ref{fg:gptspaces}(a) and Fig.~\ref{fg:gptspaces}(c), it is clear that the GPTs representing both quantum theory and Boxworld are consistent with this data. The GPT for a classical 4-level system (i.e. the $k=4$ generalization of the classical bit in Fig.~\ref{fg:gptspaces}(b)~\cite{janotta14}) is as well.

When there is a large gap between $\mathcal{S}_{\rm realized}$ and $\mathcal{S}_{\rm consistent}$, it is important to
consider the possibility that this is due to a shortcoming in the experiment and that probing more experimental configurations will reduce it. 
For instance, if an experiment on a 2-level system was governed by quantum theory, but the experimenter only considered experimental configurations involving eigenstates of Pauli operators, then  $\mathcal{S}_{\rm realized}$ and $\mathcal{E}_{\rm realized}$ would be precisely those of the example we have just described (depicted in Fig.~\ref{fg:gptspaces}(d)), implying many possibilities besides quantum theory for the true GPT.  However, further experimentation would reveal that this seemingly large scope for deviations from quantum theory was merely an artifact of probing a too-sparse set of configurations.
Only if one continually fails to close the gap between $\mathcal{S}_{\rm realized}$ and $\mathcal{S}_{\rm consistent}$, in spite of probing the greatest possible variety of experimental configurations, should one consider the possibility that in fact $\mathcal{S}\simeq \mathcal{S}_{\rm realized}$ and $\mathcal{E} \simeq \mathcal{E}_{\rm realized}$ and that the true GPT fails to satisfy the no-restriction hypothesis. By contrast, if the gap between $\mathcal{S}_{\rm realized}$ and $\mathcal{S}_{\rm consistent}$ is very small, the experiment has found a tightly constrained range of possibilities for the true GPT, and it successfully rules out a large class of alternative theories.

\section{Self-consistent tomography in the GPT framework}

We have just seen that any real experiment defines a set of realized GPT states, $\mathcal{S}_{\rm realized}$, and a set of realized GPT effects, $\mathcal{E}_{\rm realized}$, and it is from these that one can infer the scope of possibilities for the true spaces, $\mathcal{S}$ and $\mathcal{E}$, and thus the scope of possibilities for deviations from quantum theory.

\begin{figure}
\includegraphics[width=0.34\textwidth]{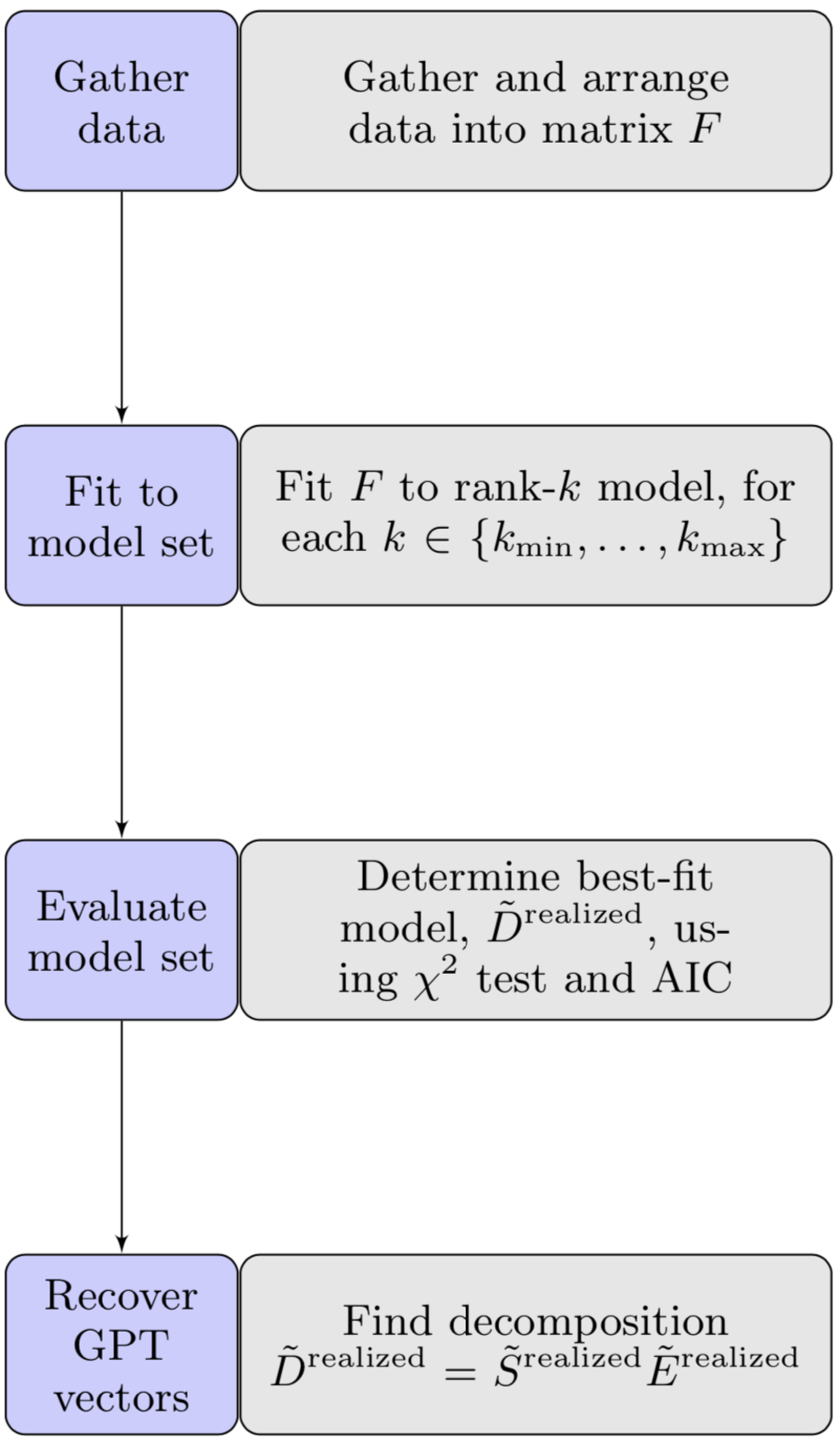}
\caption{(Color). Overview of the self-consistent GPT tomography procedure. We begin with the experimental data, finite-run relative frequencies for each configuration realized in the experiment, and arrange it into a matrix, $F$, which is a noisy version of the matrix of true probabilities, $D^{\rm realized}$. 
To estimate the dimension, $k$, of the data, we find the rank-$k$ matrix which best fits $F$ for a set of values of $k$. We call this set of best-fit rank-$k$ matrices the {\em candidate model set}. A statistical analysis on the candidate model set (using the $\chi^2$ goodness-of-fit test and the Akaike information criterion) determines the value of $k$ that gives us the best fit, and therefore which of the candidate models is the best approximation to $D^{\rm realized}$. We denote this best approximation by $\tilde{D}^{\rm realized}$. We find a decomposition $\tilde{D}^{\rm realized}=\tilde{S}^{\rm realized}\tilde{E}^{\rm realized}$, in order to estimate the spaces of states and effects realized in the experiment.
Each row of $\tilde{S}^{\rm realized}$ is a GPT state vector representing one of the preparation procedures in the experiment, and each column of $\tilde{E}^{\rm realized}$ is a GPT effect vector representing one of the measurement procedures. This completes the GPT tomography procedure.}  
\label{flowchart}
\end{figure}

But how can one estimate $\mathcal{S}_{\rm realized}$ and $\mathcal{E}_{\rm realized}$ from experimental data?  In other words, how can one implement tomography within the GPT framework?  This is the problem whose solution we now describe.
The steps in our scheme are outlined in Fig.~\ref{flowchart}.

\subsection{Tomographic completeness and the precision strategy for discovering dimensional deviations}\label{sc:tomocomplevidence}

In the introduction, we distinguished two ways in which the true GPT describing a given degree of freedom might deviate from quantum expectations.  The first possibility for deviations was in the {\em shapes} of the state and effect spaces, assuming no deviation in the dimension of the GPT vector space in which these are embedded.  The second possibility was more radical---a deviation in the dimension.  In this section, we evaluate what sort of evidence one can obtain about the dimension of GPT required to model a given degree of freedom.

We presume that there is a principle of individuation for different degrees of freedom, which is to say a way to distinguish what degree of freedom an experiment is probing.  For instance, we presume that we can identify certain experimental operations as preparations and measurements {\em of photon polarization} and not of some other degree of freedom.  

As noted earlier, the dimension of the GPT vector space associated to a degree of freedom is the minimum cardinality of a tomographically complete set of preparations (or measurements) for that degree of freedom.
Therefore, for the dimension implied by our data analysis to be the true dimension,  the sets of preparations and measurements that are experimentally realized must be tomographically complete for that degree of freedom.

Because one cannot presume the correctness of quantum theory, however, one does not have any theoretical grounds for deciding which sets of measurements (preparations) are tomographically complete for a given system. 
Indeed, whatever set of preparations (measurements) one considers as a candidate for a tomographically complete set, one can never rule out the possibility that tomorrow a novel variety of preparations (measurements) will be identified whose statistics are {\em not} predicted by those in the putative tomographically complete set, thereby demonstrating that the set was not tomographically complete after all.  As such, any supposition of tomographic completeness is always tentative.  

As Popper emphasized, however, {\em all} scientific claims are vulnerable to being falsified and therefore have a tentative status~\cite{Popper}. We are therefore recommending to treat the hypothesis that a given set of measurements and a given set of preparations are tomographically complete as Popper recommends treating any scientific hypothesis: one should try one's best to falsify it and as long as one fails to do so, the hypothesis stands. 

As noted in the introduction, it is useful to distinguish between two types of opportunities for 
falsifying a hypothesis about what sets of preparations and measurements are tomographically complete:
  terra nova strategies and precision strategies.  In this article, we pursue the latter approach.
To explain how a precision strategy provides an opportunity for detecting deviations from the quantum prediction for the dimension of the GPT vector space,
we offer an illustrative analogy.

Suppose that the GPT describing the world is indeed quantum theory.  Now consider an experiment on photon polarization wherein the experimentally realized preparations and measurements are restricted to a real-amplitude subalgebra of the full qubit alebra, that is, a {\em rebit} subalgebra. 
  In this case, the realized GPT state and effects correspond, respectively, to a restriction of the Bloch ball in Fig.~\ref{fg:gptspaces}(a)(i) to an equatorial disc and to a restriction of the ball-based diamond in Fig.~\ref{fg:gptspaces}(a)(ii)-(iii) to the diamond with the disc as its base (which is the 3-dimensional projection, depicted in Fig.~\ref{fg:gptspaces}(a)(ii), of the full 4-dimensional qubit effect space).

Suppose an experimenter did not know the ground truth about the GPT describing photon polarization, which by assumption in our example is the GPT associated to the full qubit algebra.  If they mistakenly presumed that the preparations and measurements realized in the rebit experiment  were tomographically complete, they would be led to a false conclusion about the GPT describing photon polarization.  
Nonetheless, and this is the point we wish to emphasize, high-precision experimental data provides them with an opportunity for recognizing their mistake.

The key observation is that the only case in which the experimental data contains strictly {\em no} evidence of states and effects beyond the restricted subalgebras is if the realized preparations and measurements obey the restriction {\em exactly}.
However, any real implementation of experimental procedures is necessarily imperfect, and certain types of imperfections (e.g., systematic errors) will result in preparations and measurements that {\em do} extend into the higher-dimensional space---in our example, 
from the rebit spaces into the full qubit spaces, hence from dimension 3 to dimension 4.  For instance, they might lead to preparations that were not strictly restricted to an equatorial disc but rather a fattened pancake-shaped subset of the Bloch ball, and similarly for the measurements.   The realized preparations and measurements in this case would still be very far from sampling the full qubit state and effect spaces,  but they would nonetheless attest to the need for a GPT vector space of dimension 4 rather than one of dimension 3.  Of course, if the deviation is small, then one requires a correspondingly small degree of statistical error in the characterization of the state and effect spaces  in order to detect it.   Hence the need for precision in the characterization of the states and effects.

If, in our imagined example, an experimentalist detected a deviation from their expectations regarding dimensionality in this fashion, they would be prompted 
 to look for new preparations and measurements that might extend further into this 4th dimension.  We can easily imagine that, via such a precision-based discovery of an anomaly, the experimentalist could come to learn that what at first appeared to be a rebit was in fact a qubit.

We can now draw the analogy between this sort of example and the experiment we analyze here.  
Despite the fact that we did not intentionally seek to do anything exotic in our preparations and measurements of photon polarization, it could nonetheless be the case that the GPT vectors representing these 
 had small components in additional dimensions of GPT vector space, beyond the 4 dimensions that quantum theory stipulates as sufficient for modelling photon polarization.  In this case, our scheme would find
 that the data is only fit well by a GPT of dimension greater than 4.  To the extent that one was confident that the experimental procedures did not inadvertently probe some additional degrees of freedom beyond photon polarization, this 
 would constitute evidence for postquantum physics.

We turn now to describing the self-consistent GPT tomography procedure.

\subsection{Inferring best-fit probabilities from finite-run statistics}

We suppose that, for a given system, the experimenter makes use of a finite number $m$ of preparation procedures ($P_i,\,i\in \{1,\cdots,m\}$) and a finite number, $n$, of binary-outcome measurement procedures ($M_j,\,j\in \{1,\cdots,n\}$). We denote the outcome of each measurement by $a \in \{0,1\}$.
 For each choice of preparation and measurement, $(P_i, M_j)$, the experimenter records the outcome of the measurement in a large number of runs and computes the relative frequency with which a given outcome $a$ occurs, denoted $f(a|P_i,M_j)$.  
For the binary-outcome measurements under consideration, it is sufficient to specify $f(0|P_i,M_j)$ for each pair $(P_i, M_j)$, because $f(1|P_i,M_j)$ is then fixed by normalization.

 The set of all experimental data, therefore, can be encoded in an $m\times n$ matrix $F$, whose $(i,j)$th component is $f(0|P_i,M_j)$.  
 
The relative frequency $f(0|P_i,M_j)$ one measures 
will not coincide exactly with the probability $p(0|P_i,M_j)$ from which it is assumed that the outcome in each run is sampled.\footnote{Note that it is presumed that the outcome variables for the different runs (on a given choice of preparation and measurement) are identically and independently distributed.  This assumption could fail, for instance, due to a drift in the nature of the preparation or measurement over the timescale on which the different runs take place, or due to a memory effect that makes the outcomes in different runs correlated.  In such cases, one would require a more sophisticated analysis than the one described here.} 
 Rather, $f(0|P_i,M_j)$ is merely a noisy approximation to $p(0|P_i,M_j)$.  
The statistical variation in $f(0|P_i,M_j)$ can, however, be estimated from the experiment.
 
It follows that the matrix $F$ extracted from the experimental data is merely a noisy approximation to the matrix $D^{\rm realized}$ that encodes the predictions of the GPT for the $mn$ experimental configurations of interest.
  Because of the noise, $F$ will generically  be full rank, regardless of the rank of $D^{\rm realized}$~\cite{feng07}. Therefore, 
  the experimentalist is tasked with estimating the $m\times n$ probability matrix $D^{\rm realized}$ given the $m\times n$ data matrix $F$, where the rank of $D^{\rm realized}$ is a parameter in the fit.

We aim to describe our technique in a general manner, so that it can be applied to any experiment. 
 However, in order to provide a concrete example of its use,  we will intersperse our presentation of the technique with details about how it is applied to 
  the particular experiment we have conducted.  We begin, therefore, by providing the details of the latter.
 
\subsection{Description of the experiment}

\begin{figure}
  \centering
  \includegraphics[width=0.48\textwidth]{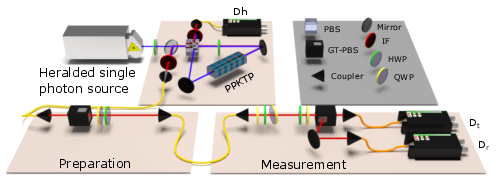}
  \caption{(Color). The experimental setup. Pairs of polarization-separable single photons are created via spontaneous parametric down-conversion. The herald photon is sent to a detector. The signal photon's polarization travels through a polarizer then a quarter and half waveplate, which prepares its polarization state. The photon is then coupled into single-mode fibre which removes any information which may be encoded in the photon's spatial degree-of-freedom. Three static waveplates undo the polarization rotation caused by the fibre. Two waveplates and a polarizing beamsplitter with detectors in each output port perform a measurement on the photon. One output port is labelled `0', and the other is labelled `1'. Coincident detections between the herald detector, $D_h$, and the detector in the transmitted port, $D_t$, are counted, as well as coincidences between $D_h$ and the reflected-port detector $D_r$.
    PPKTP: Periodically-poled potassium titanyl phosphate; PBS: Polarizing beamsplitter; GT-PBS: Glan-Thompson polarizing beamsplitter; IF: Interference filter; HWP: Half waveplate; QWP: Quarter waveplate.}
  \label{fg:setup}
\end{figure}

To illustrate the GPT tomography scheme, we perform an experiment on the polarization degree of freedom of single photons. Pairs of photons are created via spontaneous parametric down-conversion, and the detection of one of these photons, called the herald, indicates the successful preparation of the other, called the signal. We manipulate the polarization of the signal photons with a quarter- and half- waveplate before they are coupled into a single-mode fibre; each preparation is labelled by the angles of these two waveplates.

Upon emerging from the fibre, the signal photons encounter the measurement stage of the experiment, which consists of a quarter- and half-waveplate followed by a polarizing beam splitter with single-photon detectors at each of its output ports. Each measurement is labelled by the angles of the two waveplates preceding the beam splitter.

The frequency of the $0$ outcome is defined as the ratio of the number of heralded signal photon detections in the $0$ output port to the total number of heralded detections. We ignore experimental trials in which either the herald or the signal photon is lost by post-selecting on coincident detections, so that our measurements are only performed on normalized states. This is akin to making a {\em fair-sampling assumption}, that is, we assume that the statistics of the detected photons are representative of the statistics we would have measured if our experiment had perfect efficiency. Post-selecting on coincident detections has the additional benefit of allowing us to filter out background counts that are caused by, for example, stray room light or ``dark'' counts from our detectors.

We choose $m=100$ waveplate settings for the preparations, and $n=100$ waveplate settings for the settings, corresponding to $mn= 10^4$ experimental configurations in all, one for each pairing.

We choose $m=n$ so that the GPT state space and the GPT effect space are equally well characterized. We detect coincidences at a rate of $\sim2250$ counts/second, and count coincidences for each preparation-measurement pair for a total of eight seconds, allowing us  to achieve a standard deviation on each data point below the $1\%$ level.  Because of the additional time it takes to mechanically rotate the preparation and measurement waveplates, it takes approximately 84 hours to acquire data for $10^4$ preparation-measurement pairs.
 
Our method of selecting {\em which} 100 waveplate settings to use is described in Appendix~\ref{sc:targetstates}.  Note that although the choice of these settings is motivated by our knowledge of the quantum formalism, our tomographic scheme does not assume the correctness of quantum theory: our reconstruction scheme could have been applied equally well if the waveplate settings had been chosen at random.\footnote{An interesting question for future research is how the quality of the GPT reconstruction varies with the particular set of waveplate settings that are considered. In particular, one can ask about the quality of the evidence for quantum theory  in the situation wherein the waveplate settings correspond to sampling highly {\em nonuniformly} over the points on the Bloch sphere.}

\begin{figure}
  \centering
  \includegraphics[width=0.5\textwidth]{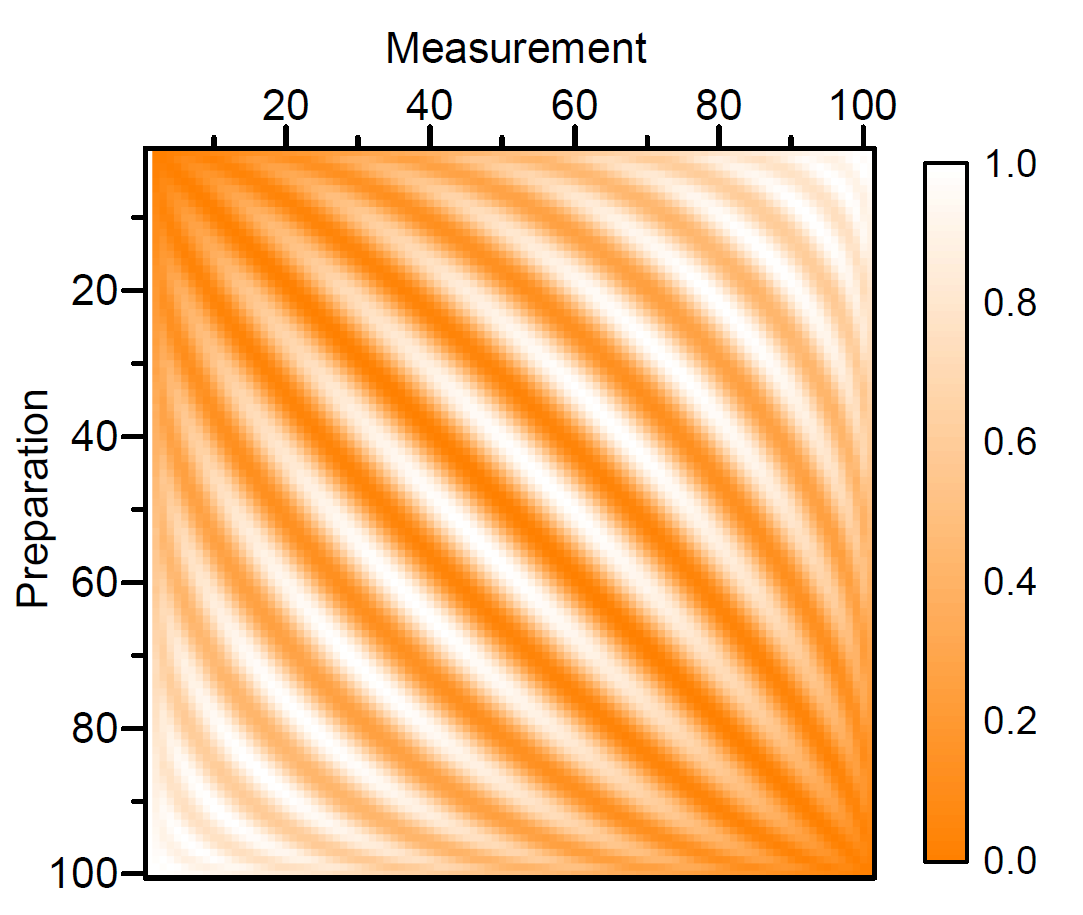}
  \caption{(Color). The raw frequencies at which outcome $a=0$ was obtained for every pair of preparation and measurement settings. The maximum standard deviations in the data are $\sim 4\times10^{-3}$. Every entry in the left-most column 
  is equal to 1---this represents the unit measurement effect which returns $a=0$ regardless of the state of the input. The striped pattern of the data is simply an artefact of the order in which we chose to implement the preparations and measurements (described in App.~\ref{sc:targetstates}).}
  \label{fg:data}
\end{figure}

The raw frequencies are arranged into the data matrix $F$. Entry $F_{ij}$ is the frequency at which the 0 outcome was obtained when measurement $M_j$ was performed on a photon that was subjected to preparation  $P_i$.
As noted in Sec.~\ref{basics}, we adopt a convention wherein $M_1$ is the unit measurement, implying that the first column of $F$ is a column of $1$s. 
The data matrix for our experiment is presented in Fig.~\ref{fg:data}. As expected, we find that $F$ is full rank.  

\subsection{Estimating the probability matrix $D^{\rm realized}$}\label{sc:estimatingDrealized}

We turn now to the problem of estimating from $F$ the $m\times n$ probability matrix $D^{\rm realized}$.
The first item of business is to estimate the rank of $D^{\rm realized}$, which is equivalent to estimating the cardinality of the tomographically complete set of preparations (or measurements) of the GPT model of the experiment.

For a given hypothesis $k$ about the value of the rank, and for a given data matrix $F$, we find the rank-$k$ matrix $\tilde{D}^{\rm realized}$ that is the maximum likelihood estimate of the rank-$k$ probability matrix $D^{\rm realized}$ that generated $F$. In other words, $\tilde{D}^{\rm realized}$ is the rank-$k$ matrix that minimizes the weighted $\chi^2$ statistic, defined as $\chi^2 \equiv \sum\limits_{i} \sum\limits_{j} \frac{\left(\tilde{D}^{\rm realized}_{ij} - F_{ij}\right)^2}{\left(\Delta F_{ij}\right)^2}$, where $\left(\Delta F_{ij}\right)^2$ is the statistical variance in $F_{ij}$.  This minimization problem is known as the weighted low-rank approximation problem, which is a non-convex optimization problem with no analytical solution~\cite{gillis11,markovsky12}.  Nonetheless, one can use a fitting algorithm based on an alternating-least-squares method~\cite{markovsky12}.  In the algorithm, it is important to constrain the entries of $\tilde{D}^{\rm realized}$ to lie within the interval $[0,1]$ so that they may be interpreted as probabilities. Full details are provided in Appendix~\ref{ap:fitting}.

To estimate the rank of the true model underlying the data, one must compare different candidate model ranks.  (For our experiment, we consider  $k\in \{2,3,\dots,10\}$.)   For each candidate rank $k$, one first computes the $\chi^2$ of the maximum-likelihood model of that rank, denoted $\chi^2_k$, in order to determine the extent to which each model might {\em underfit} the data.  Second, one computes for the max-likelihood
model of each rank the Akaike  information criterion (AIC) score~\cite{akaike73,akaike74}  in order to determine the relative extent to which the various models either underfit or {\em overfit} the data.

We begin by describing the method by which one finds the rank-$k$ probability matrix $\tilde{D}^{\rm realized}$ which minimizes $\chi^2$. Note that an $m \times n$ matrix with rank $k$ is specified by a set of $r_k = k(m+n-k)$ real parameters~\cite{candes10}, thus if the true probability matrix $D^{\rm realized}$ is rank $k$, then we expect that $\chi^2_k$ will be sampled from a $\chi^2$ distribution with $mn-k(m+n-k) = (m-k)(n-k)$ degrees of freedom~\cite{numrec}. 

\begin{figure*}[htb]
  \centering
  \includegraphics[]{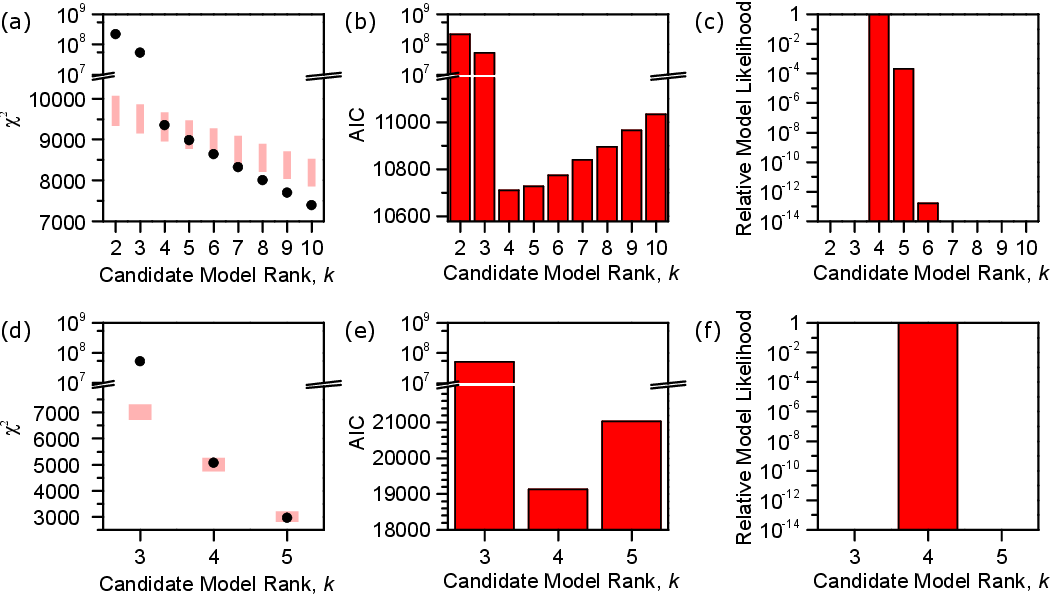}
  \caption{(Color).   Determining the true rank of the model underlying the datasets for our two experiments. (a-c) is data for the first experiment, in which we characterized 100 preparation and measurement procedures. (d-f) is for the second experiment, in which we characterized 1006 preparation and measurement procedures. (a),(d) Comparison of the fitted $\chi^2$ value to the expected value for a good fit, for various model ranks. Black circles are $\chi^2$ values returned by our fitting routine. Light red bars indicate the interval in which we expect (with 99\% confidence) the $\chi^2$-value to lie, under the assumption that the true model underlying the data is rank-$k$. Models with $k < 4$ do not fit either dataset well. (b),(e) AIC scores for each candidate model of best fit. For both datasets the rank-4 model has the lowest AIC score, and therefore is most likely the best model among the set of candidate models. (c),(f) Relative likelihood of each model in the set of candidate models (each model without a bar has a relative likelihood less than $10^{-25}$). For both datasets, the rank-4 model is most likely to describe the data.}
  \label{fg:rank}
\end{figure*}

For our experiment, we calculate the variances $(\Delta F_{ij})^2$ in the expression for $\chi^2$ by assuming that the number of detected coincident photons follows a Poissonian distribution.  Fig.~\ref{fg:rank}(a) displays the interval containing 99\% of the probability density for a $\chi^2$ distribution with $(m-k)(n-k)$ degrees of freedom, as well as $\chi^2_k$, for each value of $k\in \{2,3,\dots,10\}$. For $k<4$, $\chi^2_k$ lies far outside the expected 99\% range, and we rule out these models with high confidence.

The Akaike information criterion assigns a score to each model in a candidate set, termed its AIC score.
The Kullback-Leibler (KL) divergence is a measure of the information lost when some probability distribution $f$ is used to represent some other distribution $g$~\cite{burnhamanderson2002}, and the AIC score of a candidate model is a measure of the Kullback-Leibler (KL) divergence between the candidate model and the true model underlying the data. Since the true model isn't known, the KL divergence can't be calculated exactly. What each candidate model's AIC score represents is its KL divergence from the true model, {\em relative} to all models in the candidate set. The candidate model with the lowest AIC score is closest to the true model (in the KL sense), and thus it is the most likely representation of the data among the set of candidates.

The  AIC scores can be used to determine which model among a set of candidate models is the most likely to describe the data.  
If ${\rm AIC}_{k}$ denotes the AIC score of the $k$th model, and $\Delta_k$ denotes the difference between this score and the minimum score among all candidate models, $\Delta_k := {\rm AIC}_k - {\rm min}_{k'}{\rm AIC}_{k'}$, then its AIC {\em weight} is defined as $w_k := {e^{-\frac{1}{2}\Delta_k}}/{\sum\limits_{k=2}^{10} e^{-\frac{1}{2}\Delta_k}}$~\cite{burnhamanderson2002}.
 The AIC weight $w_k$ represents the likelihood that the $k$th model is the model that best describes the data, relative to the other models in the set of candidate models.

In our experiment, the candidate models differ by rank, and the AIC score of
a rank-$k$ candidate model is defined as ${\rm AIC}_k = \chi^2_k + 2r_k$~\cite{burnhamanderson2002}.  The first term rewards models in proportion to how well they fit the data, and the second term penalizes models in proportion to their complexity, as measured by the number of parameters.
 For our experiment, the set of candidate models is the set of best-fit rank-$k$ models for $k \in \{2,\dots,10\}$.  We plot the AIC values for each candidate model in Fig.~\ref{fg:rank}(b). $AIC_k$ is minimized for $k=4$, and we conclude that the true model underlying our dataset is most likely rank 4. 
 The relative likelihood of each candidate model is shown in Fig.~\ref{fg:rank}(c). We find $w_4 = 0.9998, w_5 = 1.99 \times 10^{-4}$, and $w_{k} < 10^{-12}$ for other values of $k$.

The $\chi^2$ goodness-of-fit test indicates that the max-likelihood rank-4 model fits the data well, and the AIC test indicates that this same model is the most likely of all nine candidate models to have generated the data, 
with relative probability 0.9998. We conclude with high confidence that the GPT that best describes our experiment has  dimension 4.

 Recall that it is still possible that the true GPT describing photon polarization has dimension greater than 4 because it is possible that the sets of preparations and measurements we have implemented in our experiment are not tomographically complete for photon polarization.

Nonetheless, the focus of much of the rest of this section and the focus of all of section~\ref{Bounding} will be to describe what additional conclusions can be drawn from our experimental data if we adopt the hypothesis that the preparations and measurements we realized are, in fact, tomographically complete for photon polarization, with the understanding that this hypothesis could in principle be overturned by future experiments that achieved higher precision or realized an exotic new variety of preparations and measurements for photon polarization.  These additional conclusions concern the possibility of deviations from quantum theory in the {\em shape} of the state and effect spaces, rather than in the dimension of the vector space in which these are embedded.

\subsection{Estimating the realized GPT state and effect spaces}\label{sc:estimating}

The realized GPT state space, $\mathcal{S}_{\rm realized}$ and the realized GPT state space, $\mathcal{E}_{\rm realized}$ define the probability matrix $D^{\rm realized}$ from which the measurement outcomes in the experiment are sampled.

 As noted above, the matrix $\tilde{D}^{\rm realized}$ for the rank-4 fit provides
our best estimate of the true probability matrix $D^{\rm realized}$. To obtain an estimate of the realized GPT state and effect spaces from $\tilde{D}^{\rm realized}$, we must decompose it in the manner described in Sec.~\ref{basics}, that is, as
 $\tilde{D}^{\rm realized} = \tilde{S}^{\rm realized} \tilde{E}^{\rm realized}$.

Recall that this decomposition is not unique.
A convenient choice is a modified form of the singular-value decomposition, where one constrains the first column of $\tilde{S}^{\rm realized}$ to be a column of ones, and one constrains the other columns of $\tilde{S}^{\rm realized}$ to be orthogonal to the first (a detailed description of this decomposition is given in Appendix~\ref{ap:decomp}). 

If quantum theory is the correct theory of nature, then the experimental data should be consistent with the GPT state space being the Bloch Ball and the GPT effect space being the Bloch Diamond (depicted in Fig.~\ref{fg:gptspaces}(a)), up to a linear invertible transformation. 

Our estimate of the realized GPT state space, $\tilde{\mathcal{S}}_{\rm realized}$, is simply the convex hull of the 
rows of the matrix $\tilde{S}^{\rm realized}$. In the case of the effects, we can again take convex mixtures, but because one also has
the freedom to post-process measurement outcomes, our estimate of the realized GPT effect space is slightly more complicated. 

There are two classes of convexly extremal classical post-processings that can be performed on a binary-outcome measurement. We call the first class of convexly extremal post-processings the {\em outcome-swapping} class.
In such a post-processing, the outcome returned by a measurement device is deterministically swapped to the other outcome. The outcome-swapping of the
outcome-0 effect for a specific measurement procedure, ${\bf e}_{[0|M]}$, is represented by that measurement's outcome-1 effect, ${\bf e}_{[1|M]}$, 
which is the complement of $\mathbf{e}_{[0,M]}$ relative to the unit effect, ${\bf e}_{[1|M]}:= \mathbf{u}-\mathbf{e}_{[0,M]}$.

We call the second class of convexly extremal post-processings the {\em outcome-fixing class}. 
In such a post-processing,
the outcome returned by a measurement device is ignored, and deterministically replaced by a fixed outcome, $0$ or $1$.  For the case where the outcome is replaced by $0$, the image of this post-processing is the unit effect ${\bf u}$, and for the case where it is replaced by $1$, the image is the complement of the unit effect (represented by the zero vector).

The full set of post-processings is obtained by taking all convex mixtures of these extremal ones. 
Hence $\mathcal{\tilde{E}}^{\rm realized}$ is the closure under convex mixtures and classical post-processing of the vectors defined by the columns of the matrix $\tilde{E}^{\rm realized}$.
As we have already included the unit measurement effect in $\tilde{D}^{\rm realized}$, it is represented in $\tilde{E}^{\rm realized}$ as well. 
Therefore, $\tilde{\mathcal{E}}_{\rm realized}$ is the convex hull of the union of the set of column vectors in the matrix $\tilde{E}^{\rm realized}$ and the set of their complements.

\begin{figure*}[htb]
  \centering
  \includegraphics[]{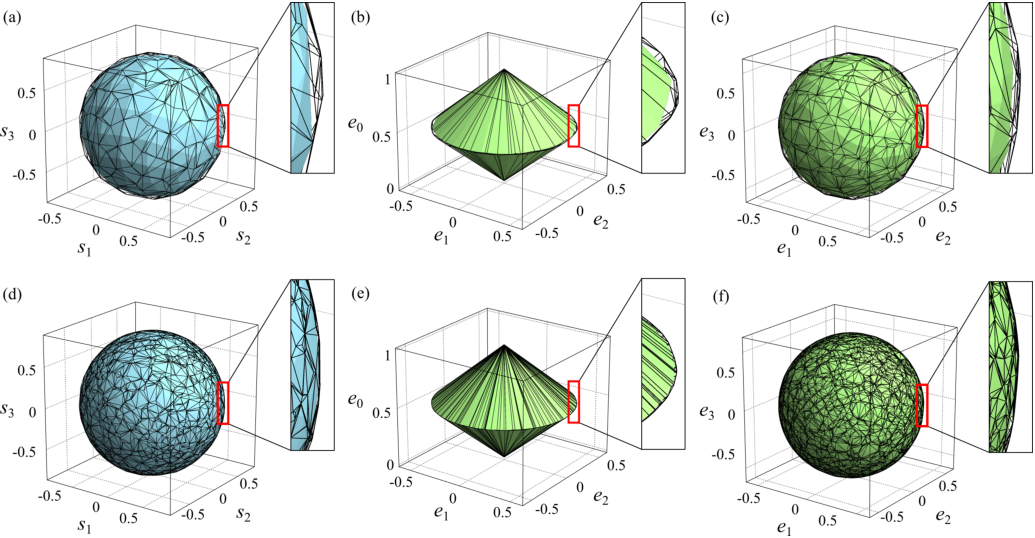}		
  \caption{(Color). 
  The GPT states and effects 
 for the preparations and measurements realized in our two experiments
   and their duals. (a),(b),(c) First experiment, in which we characterize 100 preparation and 100 measurement procedures. (d),(e),(f) Second experiment, in which we characterize 1006 preparation and 1006 measurement procedures. (a),(d) For each experiment, the estimated space of realized GPT states, $\tilde{\mathcal{S}}_{\rm realized}$ is the convex polytope depicted in blue, while the wireframe convex polytope which surrounds it is the estimated space of logically possible GPT states, $\tilde{\mathcal{S}}_{\rm consistent}$, calculated from the realized GPT effects.  The true state space of the GPT describing nature must lie somewhere in between $\tilde{\mathcal{S}}_{\rm realized}$ and $\tilde{\mathcal{S}}_{\rm consistent}$, modulo experimental uncertainty. The gap between these two spaces is smaller for the second set of data, and hence this dataset does a better job at narrowing down the possibilities for the state space. (b),(e),(c),(f) Solid green shapes are each a different 3-d projection of our estimates of the 4-d realized effect spaces, $\tilde{\mathcal{E}}_{\rm realized}$. The wireframe convex polytopes are 3-d projections of the estimated effect space consistent with the realized preparations, $\tilde{\mathcal{E}}_{\rm consistent}$.
}
  \label{fg:spaces}
\end{figure*}

Our estimate of the realized GPT state space, $\mathcal{\tilde{S}}_{\rm realized}$, and our estimate of the realized GPT effect space, $\mathcal{\tilde{E}}_{\rm realized}$, are displayed in Fig.~\ref{fg:spaces}(a)-(c). 
Omitting the first column of $\tilde{S}^{\rm realized}$ (because it contains no information), we visualize the realized GPT state space by plotting the  convex hull of the \color{black} vectors defined by the last three entries of 
 each row of $\tilde{S}^{\rm realized}$ in a 3-dimensional space (the solid light blue polytope in Fig.~\ref{fg:spaces}(a)). 
 As all four  entries of each column 
  of $\tilde{E}^{\rm realized}$ contain information, the convex hull of the vectors defined by these is 4-dimensional. To visualize the realized GPT effect space, therefore, we plot 
    two 3-dimensional projections of it, namely, the projections ${\bf e} \mapsto (e_0, e_1, e_2)$ and  ${\bf e} \mapsto (e_1, e_2, e_3)$ 
 (the solid light green polytopes in Figs.~\ref{fg:spaces}(b) and ~\ref{fg:spaces}(c) respectively).\footnote{This is the same pair of projections used to visualize the 4-dimensional GPT effect spaces depicted in Fig.~\ref{fg:gptspaces}.} 
  Qualitatively, 
 $\mathcal{S}_{\rm realized}$ is a ball-shaped polytope, and 
 $\mathcal{\tilde{E}}_{\rm realized}$ is a four-dimensional diamond
 with a ball-shaped polytope as its base.
 Note that they are qualitatively what one would expect if quantum theory is the correct description of nature.

Next, we compute the duals of these spaces.  How this is done is described in detail in Appendix~\ref{sc:dualcalc}.  Our estimate of the set of GPT state vectors that are consistent with the realized GPT effects, $\mathcal{\tilde{S}}_{\rm consistent}= {\rm dual}(\mathcal{\tilde{E}}_{\rm realized})$, is plotted alongside $\mathcal{\tilde{S}}_{\rm realized}$ in Fig.~\ref{fg:spaces}(a) as a wireframe polytope.  Similarly, our estimate of the set of GPT effect vectors consistent with the realized GPT states, $\mathcal{\tilde{E}}_{\rm consistent}= {\rm dual}(\mathcal{\tilde{S}}_{\rm realized})$, is plotted as a wireframe alongside $\mathcal{\tilde{E}}_{\rm realized}$ in Fig.~\ref{fg:spaces}(b),(c).

The smallness of the gap between $\mathcal{\tilde{S}}_{\rm realized}$ and $\mathcal{\tilde{S}}_{\rm consistent}$ implies that the possibilities for the true GPT are quite limited.  
Obviously, our results easily exclude all of the nonquantum examples of GPTs presented in Fig.~\ref{fg:gptspaces}.  

Our results can be used to infer limits on the extent to which the true GPT might fail to satisfy the no-restriction hypothesis.  One way of doing so is by bounding the volume ratio of $\mathcal{S}$ to $\mathcal{S}_{\rm logical}$.
From the discussion in Sec.~\ref{ModellingExptGPTFramework}, it is clear that this is upper bounded by the volume ratio of $\mathcal{S}_{\rm realized}$ to $\mathcal{S}_{\rm consistent}$.  Given our estimates of the latter two spaces, we can compute an estimate of this ratio.  We find it to be $0.9229\pm 0.0001$. 

The error bar is the standard deviation in the volume ratio from 100 Monte Carlo simulations. We begin each simulation by simulating a set of coincidence counts. Each set of counts is found by sampling each count from a Poisson distribution with mean and variance equal to the number of photons counted in the true experiment \footnote{Since our analysis procedure includes a constrained optimization, it is difficult to apply standard error analysis techniques to determine how errors in the measured outcome frequencies affect the GPT state and effect vectors returned by the optimization step. This is why we use a Monte Carlo analysis to estimate the errors on our estimates of the realized GPT states and effects. We note that more sophisticated error-analysis methods might give us a better estimate of the true size of the errors in our experiment, however, development of such techniques is outside the scope of this work.}.
 To our knowledge, this is the first quantitative limit on the extent to which the GPT governing nature might violate the no-restriction hypothesis.

\subsection{Increasing the number of experimental configurations}

Because the vertices of the polytopes describing $\mathcal{\tilde{S}}_{\rm realized}$ in Figs.~\ref{fg:spaces}(a)-(c)  are determined by the finite set of 
 preparations and measurement effects that were implemented, 
the observed deviation from sphericity is obviously 
an artifact of an insufficiently dense set of experimental configurations, and not evidence for any lack of smoothness of the true GPT state and effect spaces.     A higher density of experimental configurations probed in both $\mathcal{\tilde{S}}_{\rm realized}$ and $\mathcal{\tilde{S}}_{\rm consistent}$ would imply a more constrained set of possibilities for $\mathcal{S}$ and $\mathcal{S}_{\rm logical}$.  For instance, with a denser set of experimental configurations, the volume ratio of $\mathcal{\tilde{S}}_{\rm realized}$ to $\mathcal{\tilde{S}}_{\rm consistent}$ would provide a tighter upper bound on the volume ratio of $\mathcal{S}$ to $\mathcal{S}_{\rm logical}$.\footnote{Indeed, if quantum theory is correct, at sufficiently high densities of configurations, the deviation of this ratio from 1 will reflect only the unavoidable noise in every state and effect that is realized experimentally. }  As such, having a much denser set of experimental configurations would allow one to put a stronger bound on possible deviations from quantum theory, and in particular on possible deviatons from the no-restriction hypothesis.

There is therefore a strong motivation to increase the number $m$ of different preparations and the number $n$ of different measurement effects that are probed in the experiment. 
It might seem at first glance that doing so is infeasible, on the grounds that it implies a significant increase in the number, $mn$, of preparation-measurement pairs that need to be implemented and thus an overwhelmingly long data-acquisition time.

However, this is not the case; one can probe more preparations and measurements by {\em not} implementing every measurement on every preparation.  The key insight is that in order to characterize the GPT state vector associated to a given preparation, one needn't find its statistics on {\em every} measurement effect in the set being considered: it suffices to find its statistics on a subset thereof, namely, any tomographically complete subset of measurement effects.  Similarly, in order to characterize the GPT effect vector associated to a given measurement effect, one need not implement it on the full set of preparations being considered, but just a tomographically complete subset thereof.  The first experiment provided evidence for the conclusion that the tomographically complete sets have cardinality 4.
  It follows that one should be able to characterize $m$ preparations and $n$ measurements  with just $4(m+n -4)$  experimental configurations, rather than $mn$. 

Despite the good evidence about the cardinality from the first experiment, we deemed it worthwhile to perform the second experiment in such a manner that the analysis of the data did not rely on any evidence drawn from the first experiment. 
 Furthermore, we were motivated to have  the second experiment provide an independent test of the hypothesis that the cardinality of the tomographically complete sets is indeed four.   Given that the closest competitors to the rank-4 model on either side were those of ranks 3 and 5, we decided to restrict our set of candidate models to those having ranks in the set $k\in \{3,4,5\}$.  In order for the experimental data to be able to reject the hypothesis of rank $k$ as a bad fit, it is necessary that one have at least $k+1$ measurements implemented on each preparation, and at least $k+1$ preparations on which each measurement is implemented; otherwise, one can trivially find a perfect fit.  To be able to assess the quality of fit for a rank-5 model, therefore, we needed to choose at least 6 measurements that are jointly tomographically complete to implement on each of the $m$ preparations and at least $6$ preparations that are jointly tomographically complete on which each of the $n$ measurements is implemented. We chose to use precisely 6 in each case, yielding a total of $6(m+n-6)$ experimental configurations.  Without exceeding the bound of $\sim 10^4$ experimental configurations being probed (implied by the data acquisition time), we were able to take $m=n=1000$ and thereby probe a factor of 10 more preparations and measurements than in the first experiment. 

We refer to the set of six measurement effects (preparations) in this second experiment as the {\em fiducial} set. 
Our choice of which six waveplate settings to use in each of the fiducial sets is described in Appendix~\ref{sc:targetstates}.  Our choice of which 1000 waveplate settings to pair with these is also described there.  Our choices are based on our expectation that the true GPT is close to quantum theory and the desire to densely sample 
 the set of all preparations and measurements.
 (Note that although our knowledge of the quantum formalism informed our choices, our analysis of the experimental data does not presume the correctness of quantum theory.)  In the end, we also implemented each of our six fiducial measurement effects on each of our six fiducial preparations, so that we had $m=n=1006$.

We also add the unit measurement effect to our set of effects. We thereby arrange our data into a 1006$\times$1007 frequency matrix $F$, with the big difference to the first experiment being that $F$ now has
 a 1000$\times$1000 submatrix of unfilled entries. 

We perform an identical analysis procedure to the one described in Sec.~\ref{sc:estimatingDrealized}: for each $k$ in the candidate set of ranks, we seek to find the rank-$k$ matrix $\tilde{D}^{\rm realized}$ of best-fit to $F$.  For the entries in the 1000$\times$1000 submatrix of $\tilde{D}^{\rm realized}$ corresponding to the unfilled entries in $F$, 
the only constraint in the fit is that each entry be in the range $[0,1]$, so that it corresponds to a probability. The results of this analysis are presented in Fig.~\ref{fg:rank}(d)-(f).  

The $\chi^2$ goodness-of-fit test (Fig.~\ref{fg:rank}(d)) rules out the rank-3 model, and therefore all models with rank less than 3 as well. Calculating the AIC scores for the maximum-likelihood rank 3, 4 and 5 models shows that the rank-4 model is the one among these that is most likely  to describe the data (Fig.~\ref{fg:rank}(e),(f)).  Indeed the relative probability of the rank 5 model is on the order of $10^{-414}$.

The reason that the likelihood of the rank 5 model is so low is because the number of parameters required to specify a rank-$k$ $m \times n$ matrix is $r_k = k(m+n-k)$, and since $m=n\sim 1000$, the rank-5 model requires $\sim 2000$ more parameters than the rank-4 model. The number of model parameters is multiplied by a factor of two in the formula for the AIC score, and the difference between $\chi^2_5$ and $\chi^2_4$ is only $\sim 2000$. This means that if the AIC score is used to calculate the likelihood of each model, the rank 5 model is $\sim e^{-2000/2}\sim 10^{-414}$ as likely as the rank 4 model.

The AIC formula we use was derived in the limit where the number of data points is much greater than the number of parameters in the model. In our second experiment the number of data points is roughly {\em equal} to the number of parameters in each model, and thus any conclusions which derive from use of the AIC formula must be taken with a grain of salt. We should instead use
a corrected form of the AIC, called $\mathrm{AIC}_\mathrm{C}$~\cite{burnhamanderson2002}. However, the formula for $\mathrm{AIC}_\mathrm{C}$ depends on the specific model being used, and to the best of our knowledge a formula has not been found for the weighted low rank approximation problem. However, every $\mathrm{AIC}_\mathrm{C}$ formula that we found for different types of models {\em increased} the amount by which models were penalized for complexity~\cite{burnhamanderson2002}. Hence we hypothesize that the proper $\mathrm{AIC}_\mathrm{C}$ formula would lead to an even smaller relative likelihood for the rank 5 model, and thus that we have strong evidence that a rank 4 model should be used to represent the second experiment. Finding the correct $\mathrm{AIC}_\mathrm{C}$ formula for the weighted low rank approximation problem is an interesting problem for future consideration.

Modulo this caveat, the second experiment corroborates the conclusion of the first experiment, namely, that   our best estimate of the dimension of the GPT governing single-photon polarization is 4.\footnote{ As noted in our discussion of the first experiment, however, such a conclusion can in principle be overturned by future experiments if the preparations and measurements that are conventional for photon polarization exclude some exotic variety (or have undetectably small components of this exotic variety) and therefore fail to be tomographically complete.}
 
We decompose the rank-4 matrix of best fit and plot our estimates of the realized state space, $\mathcal{\tilde{S}}_{\rm realized}$, and the realized effect space, $\mathcal{\tilde{E}}_{\rm realized}$,  in Fig.~\ref{fg:spaces}(d)-(f).
The realized GPT state and effect spaces reconstructed from the second experiment are smoother than those from the first, and the gap between $\mathcal{\tilde{S}}_{\rm realized}$ and $\mathcal{\tilde{S}}_{\rm consistent}$ is smaller as well. 

The volume ratio of $\mathcal{\tilde{S}}_{\rm realized}$ to $\mathcal{\tilde{S}}_{\rm consistent}$ is found to be 
$0.977 \pm 0.001$, where the error bar is calculated from 100 Monte Carlo simulations. Compared to the first experiment, this provides a tighter bound on any failure of the no-restriction hypothesis.  

\section{Bounding deviations from quantum theory in the shape of the state and effect spaces}\label{Bounding}
 
\subsection{Consistency with quantum theory}

We now check to see if the possibilities for the true GPT state and effect spaces implied by our 
experiment include  the quantum state and effect spaces.

As noted in Sec.~\ref{ModellingExptGPTFramework}, because it is in practice impossible to eliminate all noise in the experimental procedures, we expect that under the assumption that all of our realized preparations are indeed represented by quantum states, they will all be slightly impure (that is, their eigenvalues will be bounded away from $0$ and $1$).  Their GPT state vectors should therefore be strictly in the interior of the Bloch Sphere.
 Similarly, we expect such noise on all of the realized measurement effects (with the exception of the unit effect and its outcome-swapped counterpart, which are theoretical abstractions), implying that their GPT effect vectors 
will be strictly in the interior of the 4-dimensional Bloch Diamond.  This, in turn, implies that the extremal GPT state vectors in $\mathcal{S}_{\rm consistent}$ will be strictly in the exterior of 
the Bloch Sphere.
 The size of the gap between $\mathcal{S}_{\rm realized}$ and $\mathcal{S}_{\rm consistent}$, therefore, will be determined by the  amount of noise in the preparations and measurements. 

Na\"{i}vely, one might expect that for the quantum state and effect spaces for a qubit to be consistent with our experimental results, $\mathcal{S}_{\rm qubit}$ must fit geometrically between our estimates of $\mathcal{S}_{\rm realized}$ and $\mathcal{S}_{\rm consistent}$, up to a linear transformation.  That is, one might expect the condition to be that there exists a linear transformation of $\mathcal{S}_{\rm qubit}$ that fits geometrically between $\mathcal{\tilde{S}}_{\rm realized}$ and $\mathcal{\tilde{S}}_{\rm consistent}$.

However, noise in the experiment also leads to 
 statistical discrepancies between the vertices of $\mathcal{\tilde{S}}_{\rm realized}$ and those of $\mathcal{S}_{\rm realized}$, and between the vertices of $\mathcal{\tilde{E}}_{\rm realized}$ and those of $\mathcal{E}_{\rm realized}$.
This noise could lead to estimates of the realized GPT state and effect vectors being longer than the actual realized GPT state and effect vectors.
 If the estimates of any of these lie {\em outside} the qubit state and effect spaces, then
 one could find that it is impossible to find a linear transformation of  $\mathcal{S}_{\rm qubit}$ that fits between $\mathcal{\tilde{S}}_{\rm realized}$ and $\mathcal{\tilde{S}}_{\rm consistent}$, even if quantum theory is correct!

We test the above intuition by simulating the first experiment under the assumption that quantum theory is the correct theory of nature. We assume that the states we actually prepare in the lab are slightly depolarized versions of the set of 100 pure quantum states that we are targeting,
 and that the measurements we actually perform are slightly depolarized versions of the set of 100 projective measurements we are targeting. 
 We estimate the amount of depolarization noise from the raw data, and use the estimated amount of noise to calculate the outcome probabilities for each depolarized measurement on each depolarized state. We arrange these probabilities into a $100 \times 100$ table and use them to simulate 1000 sets of photon counts, then analyse each of the 1000 simulated datasets with the GPT tomography procedure.

We find that, for every set of simulated data, we are unable to find a linear transformation of $\mathcal{S}_{\rm qubit}$ that fits between the simulated $\mathcal{\tilde{S}}_{\rm realized}$ and $\mathcal{\tilde{S}}_{\rm consistent}$, confirming the intuition articulated above.

Nonetheless, we can quantify the closeness of the fit as follows.  We find that if, for each simulation, we artificially reduce the length of the GPT vectors in the simulated $\mathcal{\tilde{S}}_{\rm realized}$ and $\mathcal{\tilde{E}}_{\rm realized}$ by multiplying them by a factor slightly less than one, then we {\em can} fit a linearly transformed $\mathcal{S}_{\rm qubit}$ between the smaller $\mathcal{\tilde{S}}_{\rm realized}$ and larger $\mathcal{\tilde{S}}_{\rm consistent}$. On average, we find we have to shrink the vectors making up $\mathcal{\tilde{S}}_{\rm realized}$ and $\mathcal{\tilde{E}}_{\rm realized}$ by $0.11\% \pm 0.02\%$,
where the error bar is the standard deviation over the set of simulations. To perform the above simulations we used CVX, a software package for solving convex problems~\cite{cvx,grant08}.

We quantify the real data's agreement with the simulations by performing the same calculation as on the simulated datasets. We first notice that there is no linear transformation of $\mathcal{S}_{\rm qubit}$ that fits between $\mathcal{\tilde{S}}_{\rm realized}$ and $\mathcal{\tilde{S}}_{\rm consistent}$, as in the simulations.  Furthermore, we find that we can achieve a fit if we shrink the vectors making up $\mathcal{\tilde{S}}_{\rm realized}$ and $\mathcal{\tilde{E}}_{\rm realized}$ by $0.14\%$, which is consistent with the simulations.
Thus the spaces $\mathcal{\tilde{S}}_{\rm realized}$ and $\mathcal{\tilde{E}}_{\rm realized}$ reconstructed from the first experiment are consistent with what we expect to find given the correctness of quantum theory.

When analysing data from the second experiment it takes $\sim 4$ hours to run the code that solves the weighted low rank approximation problem. It is therefore impractical to perform 1000 simulations of this experiment. Instead, we extrapolate from the simulation of the first experiment.  

We note two significant ways in which the second experiment differs from the first. First, we perform approximately 10 times as many preparation and measurement procedures in the second experiment than in the first, yet accumulate roughly the same amount of data. Hence, each GPT state and effect vector in the second experiment is characterized with approximately 10 times fewer detected photons than in the first experiment, and so we expect the uncertainties on the second experiment's reconstructed GPT vectors to be $\sim \sqrt{10}$ times larger than the same uncertainties in the first experiment. We expect this $\sqrt{10}$ increase in uncertainty to translate to a $\sqrt{10}$ increase in the amount we need to shrink $\mathcal{\tilde{S}}_{\rm realized}$ and $\mathcal{\tilde{E}}_{\rm realized}$ before we can fit a linearly transformed $\mathcal{S}_{\rm qubit}$ between $\mathcal{\tilde{S}}_{\rm realized}$ and $\mathcal{\tilde{S}}_{\rm consistent}$. Second, $\mathcal{\tilde{S}}_{\rm realized}$ and $\mathcal{\tilde{E}}_{\rm realized}$ each contain 1006 GPT vectors, a factor of 10 more than in the first experiment. Since there are a greater number of GPT vectors in the second experiment it is likely that the outliers (i.e., the cases for which our estimate differs most from the true vectors) in the second experiment will be more extreme than those in the first experiment. This should also lead to an increase in the amount we need to shrink the vectors in $\mathcal{\tilde{S}}_{\rm realized}$ and $\mathcal{\tilde{E}}_{\rm realized}$before we can fit a linearly transformed $\mathcal{S}_{\rm qubit}$ between $\mathcal{\tilde{S}}_{\rm realized}$ and $\mathcal{\tilde{S}}_{\rm consistent}$.

We find that, for the data from the second experiment, we need to shrink $\mathcal{\tilde{S}}_{\rm realized}$ and $\mathcal{\tilde{E}}_{\rm realized}$ by $0.65\%$, a factor only 4 times greater than the $0.14\%$ of the first experiment, which seems reasonable given the estimates above. 
We therefore conclude that the second experiment gives us no compelling reason to doubt the correctness of quantum theory.

The arguments presented above also support the notion that our experimental data is consistent with quantum theory according to the usual standards by which one judges this claim: if we had considered fitting the data with quantum states and effects rather their GPT counterparts (which one could accomplish by doing a GPT fit while {\em constraining} the vertices of the realized and consistent GPT state spaces to contain a sphere between them, up to linear transformations), we would have found that the quality of the fit was good.

\subsection{Upper and lower bounds on violation of noncontextuality inequalities}

One method we use to bound possible deviations from quantum theory is to consider the maximal violation of a particular noncontextuality inequality~\cite{spekkens09}. From our data we infer a range in which the maximal violation can lie, and compare this to the quantum prediction. We will briefly introduce the notion of noncontextuality, then discuss the inferences we make.
The notion of noncontextuality was introduced by Kochen and Specker~\cite{kochen68}.  We here consider a generalization of the Kochen-Specker notion, termed universal noncontextuality, defined in Ref.~\cite{spekkens05}.

Noncontextuality is a notion that applies to an ontological model of an operational theory.  Such a model 
 is an attempt to understand the predictions of the operational theory in terms of a system that acts as a causal mediary between the preparation device and the measurement device. It postulates a space of ontic states $\Lambda$, where the ontic state $\lambda\in\Lambda$  specifies all the physical properties of the physical system according to the model. 
For each preparation procedure $P$ of a system, it is presumed that the system's ontic state $\lambda$ is sampled at random from a probability distribution $p(\lambda|P)$. 
For each measurement $M$ on a system, it is presumed that its outcome $O$ is sampled at random in a manner that depends on the ontic state $\lambda$, based on the conditional probability 
$p(O|\lambda,M)$.
It is presumed that the empirical predictions of the operational theory are reproduced by the ontological model,
\begin{equation} \label{empiricaladequacy}
p(O | M,P) =
\sum_{\lambda\in\Lambda} p(O|\lambda,M) p(\lambda|P).
\end{equation}

We can now articulate the assumption of noncontextuality for both the preparations and the measurements.
{\bf Preparation noncontextuality.} If two preparation procedures, $P$ and $P'$, are {\em operationally equivalent}, which in the GPT framework corresponds to being represented by the same GPT state vector, then they are represented by the same distribution over ontic states:
\begin{align}
\mathbf{s}_P = \mathbf{s}_{P'}  \implies p(\lambda|P) = p(\lambda|P').\label{PNC}
\end{align}
{\bf Measurement noncontextuality.} If two measurement effects, $[O|M]$ and $[O'|M']$, are {\em operationally equivalent}, which in the GPT framework corresponds to being  represented by the same GPT effect vector, then they are represented by the same distribution over ontic states:
\begin{align}
\mathbf{e}_{[O|M]} = \mathbf{e}_{[O'|M']}  \implies p(O|\lambda,M) = p(O'|\lambda,M').
\label{MNC}
\end{align}
To assume {\em universal noncontextuality} is to assume noncontextuality for all procedures, including preparations and measurements\footnote{There is also a notion of noncontextuality for transformations~\cite{spekkens05}, but we will not make use of it here.  In fact, the noncontextuality inequality we consider is one that only makes use of the assumption of noncontextuality for preparations.}.

There are now many operational inequalities for testing universal noncontextuality.  Techniques for deriving such inequalities from proofs of the Kochen-Specker theorem are presented in \cite{kunjwal15,krishna2017deriving,kunjwal17}.
In addition, there exist other proofs of the failure of universal noncontextuality that cannot be derived from the Kochen-Specker theorem. The proofs in Ref.~\cite{spekkens05} based on prepare-and-measure experiments on a single qubit are an example, and these too can be turned into inequalities testing for universal noncontextuality (as shown in
Refs.~\cite{mazurek2016experimental} and~\cite{schmid17}).
  
We here consider the simplest example of a noncontextuality inequality that can be violated by a qubit, namely the one associated to the task of 2-bit {\em parity-oblivious multiplexing} (POM), described in Ref.~\cite{spekkens09}.
Bob receives as input from a referee an integer $y$ chosen uniformly at random from $\{0,1\}$ and Alice receives a two-bit input string $(z_0,z_1)\in \{0,1\}^2$, chosen uniformly at random.
Success in the task corresponds to Bob outputting the bit $b= z_y$, that is, the $y$th bit of Alice's input.  Alice can send a system to Bob encoding information about her input, but no information about the parity of her string, $z_0 \oplus z_1$, can be transmitted to Bob.  Thus, if the referee performs any measurement on the system transmitted, he should not be able to infer anything about the parity.  The latter constraint is termed {\em parity-obliviousness}.
\footnote{Parity-oblivious multiplexing is akin to a 2-to-1 quantum random access code.  It was not introduced as a type of random access code in Ref.~\cite{spekkens09} because the latter are generally defined  as having a constraint on the potential information-carrying capacity of the system transmitted, whereas in parity-oblivious multiplexing, the system can have arbitrary information-carrying capacity---the only constraint is that of parity-obliviousness.}

An operational theory describes every protocol for parity-oblivious multiplexing as follows. Based on the input string
$(z_0,z_1)\in \{0,1\}^2$
 that she receives from the referee, Alice implements a preparation procedure $P_{z_0 z_1}$, and based on the integer $y \in \{0,1\}$ that he receives from the referee, Bob implements a binary-outcome measurement $M_y$, and reports the outcome $b$ of his measurement as his output. 
 Given that each of the 8 values of $(y, z_0, z_1)$ are equally likely, the probability of winning, denoted $\mathcal{C}$, is
\begin{equation}
\mathcal{C} \equiv \frac{1}{8} \sum_{b, y,z_0,z_1} \delta_{b,z_y} p(b |P_{z_0 z_1},M_y).
\label{defnPOMquantity}
\end{equation}
where $\delta_{b,z_y}$ is the Kronecker delta function.
The parity obliviousness condition can be expressed as a constraint on the GPT states, as 
\begin{align}
\frac{1}{2} \mathbf{s}_{P_{00}}+\frac{1}{2} \mathbf{s}_{P_{11}} = \frac{1}{2} \mathbf{s}_{P_{01}}+\frac{1}{2} \mathbf{s}_{P_{10}}.
\label{parityobliviousness}
\end{align}
This asserts the operational equivalence of the parity-0 preparation (the uniform mixture of $P_{00}$ and $P_{11}$) and the parity-1 preparation (the uniform mixture of $P_{01}$ and $P_{10}$), and therefore it implies a nontrivial constraint on the ontological model by the assumption of preparation noncontextuality (Eq.~\eqref{PNC}), namely,
\begin{align}
\frac{1}{2} p(\lambda|P_{00})+\frac{1}{2} p(\lambda|P_{11}) = \frac{1}{2} p(\lambda|P_{01})+\frac{1}{2} p(\lambda|P_{10}).
\label{ConsOfPNC}
\end{align}

It was shown in Ref.~\cite{spekkens09} that if an operational theory admits of a universally noncontextual ontological model, then the maximal value of the probability of success in parity-oblivious multiplexing is 
\begin{equation}
\mathcal{C}_ {\rm NC} \equiv \frac{3}{4}.\label{NCbound}
\end{equation}
We refer to the inequality
\begin{equation}
\mathcal{C} \le \mathcal{C}_{\rm NC}
\end{equation}
as the POM noncontextuality inequality.
\footnote{Note that an experimental test of this inequality was also reported in Ref.~\cite{spekkens09}.  
However, as noted in Ref.~\cite{mazurek2016experimental}, the experiment of Ref.~\cite{spekkens09} did not solve the problem of inexact operational inequivalences.  Although the measured deviation from exact operational equivalence was found to be small, there was at the time no theoretical account of how a given value of deviation should impact the degree of violation of the POM inequality.  As such, it was unclear what conclusions could be drawn for the possibility of noncontextuality from the violation of the POM inequality in that experiment.  }

  It was also shown in Ref.~\cite{spekkens09} that 
in operational quantum theory, the maximal value of the probability of success is 
\begin{equation}
\mathcal{C}_ {\rm Q} \equiv \frac{1}{2} + \frac{1}{2\sqrt{2}}\simeq 0.8536,\label{Qbound}
\end{equation}
which violates the POM noncontextuality inequality, thereby providing a proof of the impossibility of a noncontextual model of quantum theory and demonstrating a quantum-over-noncontextual advantage for the task of parity-oblivious multiplexing.
A set of four quantum states and two binary-outcome quantum measurements that satisfy the parity-obliviousness condition of Eq.~\eqref{parityobliviousness} and that lead to success probability $\mathcal{C}_Q$ are illustrated in Fig.~\ref{fg:innerouterapprox}.

For a given GPT state space $\mathcal{S}$ and effect space $\mathcal{E}$, we define
\begin{align}
  \mathcal{C}_{(\mathcal{S},\mathcal{E})} \equiv \max_{\substack{ \{{\bf s}_{P_{z_0 z_1}}\}\in \mathcal{S}\\\{{\bf e}_{b|M_y}\}\in\mathcal{E} }}  \frac{1}{8} \sum_{b, y,z_0,z_1} \delta_{b,z_y} {\bf s}_{P_{z_0 z_1}}\cdot {\bf e}_{b|M_y},
\label{CSE}
\end{align}
where the optimization must be done over choices of $\{{\bf s}_{P_{z_0 z_1}}\}\in \mathcal{S}$ that satisfy the parity-obliviousness constraint of Eq.~\eqref{parityobliviousness}.
If $\mathcal{S}$ and $\mathcal{E}$ are the state and effect spaces of a GPT, then ${\bf s}_{P_{z_0 z_1}}\cdot {\bf e}_{b|M_y}$ is the probability $p(b |P_{z_0 z_1},M_y)$ and
$\mathcal{C}_{(\mathcal{S},\mathcal{E})}$ has the form of Eq.~\eqref{defnPOMquantity} and defines the maximum probability of success achievable in the task of parity-oblivious multiplexing for that GPT.  (We will see below that it is also useful to consider $\mathcal{C}_{(\mathcal{S},\mathcal{E})}$ when the pair $\mathcal{S}$ and $\mathcal{E}$ {\em do not} define the state and effect spaces of a GPT.)

As discussed in Section~\ref{ModellingExptGPTFramework}, no experiment can specify $\mathcal{S}$ and $\mathcal{E}$ exactly.
 Instead, what we find is a set of possibilities for 
$(\mathcal{S},\mathcal{E})$ that are consistent with the data, and thus are candidates for the true GPT state and effect spaces. We denote this set of candidates by $\texttt{GPT}_{\rm candidates}$.
To determine the range of possible values of the POM 
noncontextuality inequality violation in this set, 
we need to determine
\begin{equation}
\mathcal{C}_{\rm min}^{} \equiv \min_{(\mathcal{S},\mathcal{E})\in \texttt{GPT}_{\rm candidates}} \mathcal{C}_{(\mathcal{S},\mathcal{E})}^{},\label{Cmin}
\end{equation}
and 
\begin{equation}
\mathcal{C}_{\rm max}^{} \equiv \max_{(\mathcal{S},\mathcal{E})\in \texttt{GPT}_{\rm candidates}} \mathcal{C}_{(\mathcal{S},\mathcal{E})}^{}.\label{Cmax}
\end{equation}

\begin{figure}
  \centering
 \includegraphics{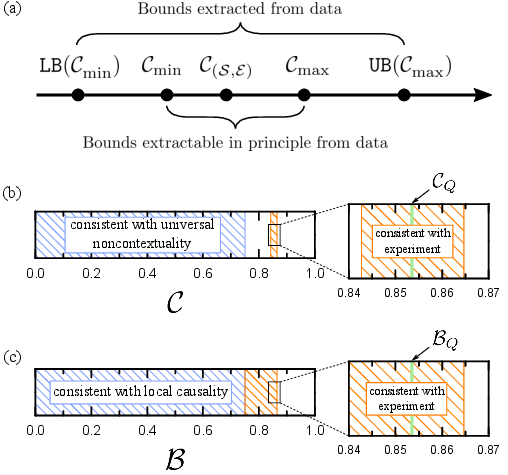}
  \caption{(Color). Bounding maximal inequality violations with GPT tomography. (a) Relation between the true value of the maximal violation of the POM inequality for the true GPT describing our experiment, $\mathcal{C}_{(\mathcal{S},\mathcal{E})}$, and the bounds that we place on it. The interval $[\mathcal{C}_{\rm min},\mathcal{C}_{\rm max}]$ is the range of possible values for $\mathcal{C}_{(\mathcal{S},\mathcal{E})}$ that one can in principle infer from an experiment, and the interval $[\texttt{LB}(\mathcal{C}_{\rm min}),\texttt{UB}(\mathcal{C}_{\rm max})]$ is a conservative estimate of $[\mathcal{C}_{\rm min},\mathcal{C}_{\rm max}]$. (b) The interval $[\texttt{LB}(\mathcal{C}_{\rm min}),\texttt{UB}(\mathcal{C}_{\rm max})]$ inferred from our data (area labelled ``consistent with experiment''). The true value $\mathcal{C}_{(\mathcal{S},\mathcal{E})}$ differs from the quantum prediction, $\mathcal{C}_Q$ by at most $\pm 1.3 \pm 0.1$\%. Our data violates the POM inequality. (c) The interval $[\texttt{LB}(\mathcal{B}_{\rm min}),\texttt{UB}(\mathcal{B}_{\rm max})]$ inferred from our data (area labelled ``consistent with experiment''). The true value $\mathcal{B}_{(\mathcal{S},\mathcal{E})}$ is at most $1.3\pm 0.1$\% greater than the maximal quantum violation, $\mathcal{C}_Q$. Error bars are too small to be visible on the plots.
}
  \label{fg:B}
\end{figure}

\begin{figure}
  \centering
  \includegraphics{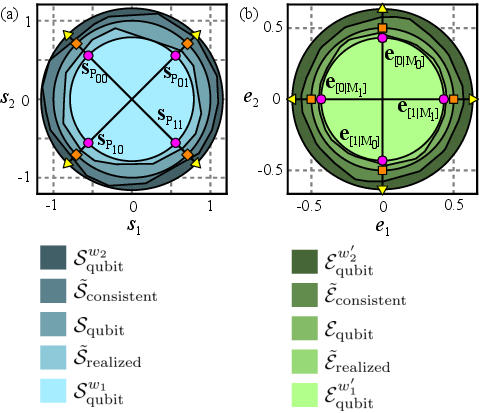}		
  \caption{(Color). Depictions of the rescaled qubit state and effect spaces which provide our inner and outer approximations to the estimated realized GPT state and effect spaces. We also depict the states and effects that achieve the maximum probability of success in parity-oblivious multiplexing in quantum theory (orange squares), and those that achieve our lower (magenta circles) and upper (yellow triangles) bounds. The left figure depicts the  GPT state vectors of the four preparations, labelled by the possible values of the pair of bits Alice must encode, and the right figure depicts the GPT effect vectors of each outcome of each of the pair of measurements. 
  }
\label{fg:innerouterapprox}
\end{figure}

See Fig.~\ref{fg:B}(a) for a schematic of the relation between the various $\mathcal{C}^{}$ quantities we consider.

$\mathcal{C}_{\rm min}$ and $\mathcal{C}_{\rm max}$ are each defined as a solution to an optimization problem. As noted in Sec.~\ref{ModellingExptGPTFramework}, there is a large freedom in the choice of $\mathcal{S}$ given $\mathcal{S}_{\rm realized}$ and $\mathcal{S}_{\rm consistent}$, and there is a large freedom in the choice of $\mathcal{E}$ for each choice of $\mathcal{S}$.  Finally, for each pair $(\mathcal{S},\mathcal{E})$ in this set, one still needs to optimize over the choice of four preparations and two measurements defining the probability of success. 

It turns out that the choice of $(\mathcal{S},\mathcal{E})$ that determines $\mathcal{C}_{\rm min}$ is easily identified.
First, note that the definition in Eq.~\eqref{CSE} implies the following inference
\begin{align}
\mathcal{S}' \subseteq \mathcal{S}, \; \mathcal{E}' \subseteq \mathcal{E} \implies \mathcal{C}_{(\mathcal{S}',\mathcal{E}')}^{} \le \mathcal{C}_{(\mathcal{S},\mathcal{E})}.
\label{inference1}
\end{align}

Given that 
$\mathcal{S}_{\rm realized} \subseteq \mathcal{S}$ and $\mathcal{E}_{\rm realized} \subseteq \mathcal{E}$ for all $(\mathcal{S},\mathcal{E})\in \texttt{GPT}_{\rm candidates}$, 
it follows that 
\begin{equation}
\mathcal{C}_{(\mathcal{S}_{\rm realized},\mathcal{E}_{\rm realized})} \le \mathcal{C}_{\rm min}.
\end{equation}
And given that $(\mathcal{S}_{\rm realized},\mathcal{E}_{\rm realized})$ is among the GPT candidates consistent with the data, we conclude that 
\begin{equation}
\mathcal{C}_{\rm min} = \mathcal{C}_{(\mathcal{S}_{\rm realized},\mathcal{E}_{\rm realized})}.
\label{Cminformula}
\end{equation}

However, calculating $\mathcal{C}_{(\mathcal{S}_{\rm realized},\mathcal{E}_{\rm realized})}$ still requires solving the optimization problem defined in Eq.~\eqref{CSE}, which is computationally difficult.

Much more tractable is the problem of determining a {\em lower bound} on $\mathcal{C}_{\rm min}^{}$, using a simple inner approximation to $\mathcal{S}_{\rm realized}$ and $\mathcal{E}_{\rm realized}$.  This is the approach we pursue here.  We will denote this lower bound by $\texttt{LB}(\mathcal{C}_{\rm min}^{})$.

Let $\mathcal{S}^{w}_{\rm qubit}$ denote the image of the qubit state space $\mathcal{S}_{\rm qubit}$ under the partially depolarizing map $\mathcal{D}_w$, defined by
\begin{equation}\label{eq:depolmap}
\mathcal{D}_w (\rho) \equiv w \rho + (1-w) \frac{1}{2} \mathbb{I}\tr(\rho),
\end{equation}
with $w\in [0,1]$.
 Similarly, let $\mathcal{E}^{w'}_{\rm qubit}$ denote the image of $\mathcal{E}_{\rm qubit}$ under $\mathcal{D}_{w'}$.  

Consider the 2-parameter family of GPTs defined by $\{(\mathcal{S}^{w}_{\rm qubit},\mathcal{E}^{w'}_{\rm qubit}): w,w'\in(0,1)\}$. These correspond to quantum theory for a qubit but with noise added to the states and to the effects.  Letting $w_1$ be the largest value of the parameter $w$ such that $\mathcal{S}^{w}_{\rm qubit} \subseteq \mathcal{S}_{\rm realized}$ and letting $w'_1$ be the largest value of the parameter $w'$ such that $\mathcal{E}^{w}_{\rm qubit} \subseteq \mathcal{E}_{\rm realized}$, then $\mathcal{S}^{w_1}_{\rm qubit}$ and $\mathcal{E}^{w'_1}_{\rm qubit}$ provide inner approximations to $\mathcal{S}_{\rm realized}$ and $\mathcal{E}_{\rm realized}$ respectively, depicted in Fig.~\ref{fg:innerouterapprox}.
From these, we get the lower bound
\begin{equation}
\texttt{LB}(\mathcal{C}_{\rm min}^{}) = \mathcal{C}_{(\mathcal{S}^{w_1}_{\rm qubit},\mathcal{E}^{w'_1}_{\rm qubit})}.\label{LBw1}
\end{equation}

A subtlety that we have avoided mentioning thus far is that the depolarized qubit state and effect spaces are only defined up to a linear transformation, so that in seeking an inner approximation, one could optimize over not only $w$ but linear transformations as well.  To simplify the analysis, however, we  took $\mathcal{S}^{w}_{\rm qubit}$ to be a sphere of radius $w$ and $\mathcal{E}^{w'}_{\rm qubit}$ to be a diamond with a base that is a sphere of radius $w'$, and we optimized over $w$ and $w'$. (Optimizing over all linear transformations would simply give us a tighter lower bound.)

For the GPT $(\mathcal{S}^{w}_{\rm qubit},\mathcal{E}^{w'}_{\rm qubit})$, a set of four preparations and two binary-outcome measurements that satisfy the parity-obliviousness condition of Eq.~\eqref{parityobliviousness} and that 
yield the maximum probability of success 
are the images, under the partially depolarizing maps $\mathcal{D}_w$ and $\mathcal{D}_{w'}$ respectively, of the optimal quantum choices.   These images are depicted in Fig.~\ref{fg:innerouterapprox}. 

For this GPT, one finds that the probability of success in parity-oblivious multiplexing is the quantum value  with probability $ww'$, and 1/2 the rest of the time, 
\begin{align}
\mathcal{C}_{(\mathcal{S}^{w}_{\rm qubit},\mathcal{E}^{w'}_{\rm qubit})}^{} &= w w' (\frac{1}{2}+\frac{1}{2\sqrt{2}}) + (1- w w') \frac{1}{2},\nonumber\\
&= \frac{1}{2} + w w' \frac{1}{2\sqrt{2}}.
\label{BPOMw}
\end{align}

From our estimates of the realized GPT state and effect spaces, $\mathcal{\tilde{S}}_{\rm realized}$ and $\mathcal{\tilde{E}}_{\rm realized}$, we obtain an estimate of $w_1$ by identifying the largest value of $w$ such that $\mathcal{S}^{w}_{\rm qubit} \subseteq \mathcal{\tilde{S}}_{\rm realized}$ and we obtain an estimate of $w'_1$ by identifying the largest value of $w'$ such that 
$\mathcal{E}^{w'}_{\rm qubit} \subseteq \mathcal{\tilde{E}}_{\rm realized}$. 

Determining these estimates from the data of the first experiment, then substituting into
Eq.~\eqref{BPOMw} and using Eq.~\eqref{LBw1}, we infer the lower bound
$\texttt{LB}(\mathcal{C}_{\rm min}^{}) = 0.8303 \pm 0.0002.$
A similar analysis for the second experiment yields an even tighter bound,
\begin{equation}
\texttt{LB}(\mathcal{C}_{\rm min}^{}) = 0.8427 \pm 0.0005.\label{numLBpomp2}
\end{equation}
This provides a lower bound on the interval of $\mathcal{C}$ values in which the true value could be found, as depicted in Fig.~\ref{fg:B}(b).\footnote{Note that it is likely that this lower bound could be improved if one supplemented the preparations and measurements that were implemented in the experiment with a set that were targeted towards achieving the largest value of $\mathcal{C}$ (according to quantum expectations).}

We now turn to $\mathcal{C}_{\rm max}$.
 Given that for all $(\mathcal{S},\mathcal{E}) \in \texttt{GPT}_{\rm candidates}$, $\mathcal{S} \subseteq \mathcal{S}_{\rm consistent}$ and $\mathcal{E} \subseteq \mathcal{E}_{\rm consistent}$, it follows from Eq.~\eqref{inference1} that $\mathcal{C}_{\rm max} \le \mathcal{C}_{(\mathcal{S}_{\rm consistent},\mathcal{E}_{\rm consistent})}$.  
 \footnote{
At this point, the analogy to the case of $\mathcal{C}_{\rm min}$ might lead one to expect that
  $\mathcal{C}_{\rm max}=\mathcal{C}_{(\mathcal{S}_{\rm consistent},\mathcal{E}_{\rm consistent})}^{}$.  However, this is incorrect
because the pair $(\mathcal{S}_{\rm consistent},\mathcal{E}_{\rm consistent})$ is {\em not} among the GPT candidates consistent with the experimental data. In fact, it does not even correspond to a valid GPT, as one can find a GPT state vector in $\mathcal{S}_{\rm consistent}$ and a GPT effect vector in $\mathcal{E}_{\rm consistent}$ with inner product outside the interval $[0,1]$, hence not defining a probability. Unfortunately, if one wants to calculate $\mathcal{C}_{\rm max}$, it seems that one must perform the difficult optimization in Eq.~\eqref{Cmax}.}
We can therefore compute an upper bound on $\mathcal{C}_{\rm max}$
 using outer approximations to $\mathcal{S}_{\rm consistent}$ and $\mathcal{E}_{\rm consistent}$. 
We choose outer approximations consisting of rescaled qubit state and effect spaces, defined as before, but where the parameter $w$ can now fall outside the interval $[0,1]$.

Letting $w_2$ be the smallest value of the parameter $w$ such that $\mathcal{S}_{\rm consistent} \subseteq   \mathcal{S}^{w}_{\rm qubit} $ and letting $w'_2$ be the smallest value of the parameter $w'$ such that $\mathcal{E}_{\rm consistent} \subseteq   \mathcal{E}^{w'}_{\rm qubit} $, then $\mathcal{S}^{w_2}_{\rm qubit}$ and $\mathcal{E}^{w'_2}_{\rm qubit}$ provide outer approximations to $\mathcal{S}_{\rm consistent}$ and $\mathcal{E}_{\rm consistent}$ respectively, and so we get an upper bound
\begin{equation}
\texttt{UB}(\mathcal{C}_{\rm max}^{}) = \mathcal{C}_{(\mathcal{S}^{w_2}_{\rm qubit},\mathcal{E}^{w'_2}_{\rm qubit})}^{}.\label{UBw2}
\end{equation}

Even though we are now allowing supernormalized state and effect vectors, via $w$ and $w'$ values outside of $[0,1]$, a simple calculation shows that $\mathcal{C}_{(\mathcal{S}^{w}_{\rm qubit},\mathcal{E}^{w'}_{\rm qubit})}$ is still given by Eq.~\eqref{BPOMw}.

Our estimates $\mathcal{\tilde{S}}_{\rm consistent}$ and $\mathcal{\tilde{E}}_{\rm consistent}$ for the state and effect spaces of the first experiment imply estimates for $w_2$ and $w'_2$ \footnote{We note that the duality relation $\mathcal{E}_{\rm consistent} = {\rm dual}(\mathcal{S}_{\rm realized})$ implies that $\mathcal{E}_{\rm qubit}^{w_2'} = {\rm dual}(\mathcal{S}_{\rm qubit}^{w_1})$ and similarly, the relation $\mathcal{S}_{\rm consistent} = {\rm dual}(\mathcal{E}_{\rm realized})$ implies $\mathcal{S}_{\rm qubit}^{w_2} = {\rm dual}(\mathcal{E}_{\rm qubit}^{w_1'})$. This in turn implies that $w_2' = \frac{1}{w_1}$ and $w_2 = \frac{1}{w'_1}$, so that $\mathcal{C}_{\rm min} = \frac{1}{2} + w_1 w'_1 \frac{1}{2\sqrt{2}}$  and  $\mathcal{C}_{\rm max} = \frac{1}{2} + \frac{1}{w_1 w'_1}\frac{1}{2\sqrt{2}}$.}
and substituting these into Eqs.~\eqref{UBw2} and \eqref{BPOMw}, we infer
$\texttt{UB}(\mathcal{C}_{\rm max}^{}) = 0.8784 \pm 0.0002.$
The same analysis on the second experiment yields
\begin{equation}
\texttt{UB}(\mathcal{C}_{\rm max}^{}) = 0.8647 \pm 0.0005.\label{numUBpomp2}
\end{equation}
This provides an upper bound on the interval of $\mathcal{C}$ values in which the true value could be found, as depicted in Fig.~\ref{fg:B}(b).

Recalling that the quantum value is $\mathcal{C}_{Q}^{} \simeq 0.8536$, it follows from Eqs.~\eqref{numLBpomp2} and \eqref{numUBpomp2} that the scope for the true GPT to differ from quantum theory in the amount of contextuality it predicts (relative to the POM inequality) is quite limited: for the true GPT, the maximum violation of the POM noncontextuality inequality can be at most $1.3\% \pm 0.1$ less than and at most $1.3\% \pm 0.1$ greater than the quantum value.

\subsection{Upper bound on violation of Bell inequalities}

Bell's theorem famously shows that a certain set of assumptions, which includes local causality, is in contradiction with the predictions of operational quantum theory~\cite{bell64}.  It is also possible to derive inequalities from these assumptions that refer only to operational quantities and thus can be tested directly experimentally. 

 The Clauser, Horne, Shimony and Holt (CHSH) inequality~\cite{clauser69} is the standard example.  
 A pair of systems are prepared together according to a preparation procedure $P^{AB}$, then one is sent to Alice and the other is sent to Bob.   At each wing of the experiment, the system is subjected to one of two binary-outcome measurements, $M^A_0$ or $M^A_1$ on Alice's side and $M^B_0$ and $M^B_1$ on Bob's side, with the choice of measurement being made uniformly at random, and where the choice at one wing is space-like separated from the registration of the outcome at the other wing.  Denoting the binary variable determining the measurement choice at Alice's (Bob's) wing by $x$ ($y$), and the outcome of Alice's (Bob's) measurement by $a$ ($b$), the operational quantity of interest, the ``Bell quantity'' for CHSH, is defined as follows (where $a,b,x,y \in \{0,1\}$, and $\oplus$ is addition modulo 2)
\begin{align}
\mathcal{B} \equiv \frac14 \sum_{a,b,x,y} \delta_{a\oplus b, xy} p(a,b|M^A_x,M^B_y,P^{AB}).
\label{defnB}
\end{align}
The maximum value that this quantity can take in a model satisfying local causality and the other assumptions of Bell's theorem is 
\begin{align}
\mathcal{B}_{\rm loc}  \equiv \frac{3}{4},
\end{align}
so that such models satisfy the CHSH inequality 
\begin{align}
\mathcal{B} \le \mathcal{B}_{\rm loc}.
\end{align}
Meanwhile, the maximum quantum value is \cite{tsirelson1980quantum} 
\begin{align}
\mathcal{B}_{\rm Q}  \equiv  \frac{1}{2} +\frac{1}{2\sqrt{2}} \simeq 0.8536.
\label{eq:tsirelson}
\end{align}

Experimental tests have exhibited a violation of the CHSH inequality~\cite{aspect82} and various loopholes for escaping this conclusion have been sealed experimentally~\cite{weihs98,rowe01,erven14,hensen15,giustina15,shalm15}. These experiments provide a lower bound on the value of the Bell quantity, 
which violates the local bound.  

It has not been previously clear, however, how to derive an {\em upper} bound on the Bell quantity. 
Doing so is necessary if one hopes to 
experimentally rule out post-quantum correlations such as the Popescu-Rohrlich box~\cite{tsirelson1980quantum,popescu1994quantum}. We here demonstrate how to do so.

First note that the probability for obtaining outcomes $a$ and $b$ given settings $x$ and $y$, which appears in Eq.~\eqref{defnB}, can be expressed in the GPT framework as 
\beq
p(a,b|M^A_x,M^B_y,P^{AB}) = {\bf s}_{P^{AB}} \cdot ({\bf e}_{a|M^{A}_x} \otimes {\bf e}_{b|M^B_y} ),
\eeq
 where 
 ${\bf s}_{P^{AB}}$ is the GPT state on the composite system $AB$ representing the preparation $P^{AB}$ (it is said to be entangled if it cannot be written as a  convex mixture of states that factorize on the vector spaces of the components
~\cite{barnum16}), and where 
${\bf e}_{a|M^{A}_x}$ (${\bf e}_{b|M^B_y}$) is the GPT effect on $A$ ($B$) representing the outcome $a$ ($b$) of measurement $M^A_x$ ($M^B_y$).  
Learning that the $M^A_x$ measurement was implemented on the preparation $P^{AB}$ and yielded the outcome $a$ can be conceived of as a preparation for system $B$, which we denote by $P^B_{a|x}$.
The GPT state representing this remote preparation, which we denote by ${\bf s}_{P^B_{a|x}}$, is defined by
\beq
p_{a|x} {\bf s}_{P^B_{a|x}} :=({\bf e}_{a|M^{A}_x}\otimes I^B)^T {\bf s}_{P^{AB}},
\eeq
where we introduce the shorthand $p_{a|x} \equiv p(a|M_x^A,P^{AB})$, and where $I^B$ represents the identity operator on system $B$.
Given this definition, one can reexpress the probability appearing in the Bell quantity as 
\beq
p(a,b|M^A_x,M^B_y,P^{AB}) = p_{a|x} {\bf s}_{P^B_{a|x}} \cdot {\bf e}_{b|M^B_y},
\eeq
which involves only GPT states and GPT effects on system $B$.  
In this case, one is conceptualizing the Bell experiment as achieving one of a set of remote preparations of the state of Bob's system---commonly referred to as ``steering''---followed by a measurement on Bob's system.

The assumption of space-like separation implies that there is no signalling between Alice and Bob, and this constrains how Bob's system can be steered. Since $p_{a|x}$ is the probability that Alice obtains outcome $a$ given that she performs measurement $M_x^A$ on the preparation $P^{AB}$, the marginal GPT state of Bob's subsystem when one does not condition on $a$ is given by $\sum\limits_a p_{a|x} \mathbf{s}_{P^B_{a|x}}$. 
The no-signalling assumption forces this marginal state to be independent of Alice's measurement choice $x$. In the CHSH scenario the no-signalling constraint is summarized with the following equation:
\begin{equation}\label{eq:nosignal}
  p_{0|0} \mathbf{s}_{P^B_{0|0}} + p_{1|0} \mathbf{s}_{P^B_{1|0}}   
  = p_{0|1} \mathbf{s}_{P^B_{0|1}} + p_{1|1} \mathbf{s}_{P^B_{1|1}}.
\end{equation}

Because we are assuming that the true GPT includes classical probability theory as a subtheory (see Sec.~\ref{basics}),  it follows that the local value, $\mathcal{B}_{\rm loc}$, is a lower limit on the range of possible values of the Bell quantity among experimentally viable candidates for the true GPT. This is a trivial lower limit. 
In order to obtain a {\em nontrivial} lower limit on this range (i.e., one greater than $\mathcal{B}_{\rm loc}$),  one would need
 to perform an experiment involving two physical systems such that one can learn which GPT states for the bipartite system are physically realizable (in particular, whether there are any entangled states that are realized) and thus which steering schemes are physically realizable. Because our experiment is on a single physical system, it cannot attest to the {\em physical realizability} of any bipartite states and hence cannot attest to the physical realizability of any particular instance of steering.

Nonetheless, our experiment {\em can} attest to the {\em logical impossibility} of particular instances of steering, namely, any instance of steering wherein the ensemble on Bob's system contains one or more GPT states {\em outside} of $\mathcal{S}_{\rm consistent}$, because such states by definition assign values outside $[0,1]$---which cannot be interpreted as probabilities---to some physically realized GPT effects (i.e., some GPT effects in $\mathcal{E}_{\rm realized}$).
This in turn implies the {\em nonexistence} of any bipartite GPT state
 (together with a GPT measurement on Alice's system) 
 which could be used to realize such an instance of steering, even though the experiment probes only a single system rather than a pair.
 
Therefore, we {\em can} use our experimental results to determine an {\em upper limit} on the range of values of the Bell quantity among experimentally viable candidates for the true GPT.

The maximum violation of the CHSH inequality achievable if Bob's system is described by a state space $\mathcal{S}$ and an effect space $\mathcal{E}$, is
\begin{equation}
\mathcal{B}_{(\mathcal{S},\mathcal{E})} 
\equiv \max_{\substack{ \{p_{a|x}\}\\\{{\bf s}_{P^B_{a|x}}\}\in \mathcal{S} \\ \{{\bf e}_{b|M^B_y}\}\in\mathcal{E} }}
\frac14 \sum_{a,b,x,y} \delta_{a\oplus b, xy} p_{a|x} {\bf s}_{P^B_{a|x}}\cdot {\bf e}_{b|M^B_y},
\label{BSE}
\end{equation}
where one varies over $\{p_{a|x}\}, \{{\bf s}_{P^B_{a|x}}\}$ that satisfy the no-signalling constraint, Eq.~\eqref{eq:nosignal}.  If the pair $\mathcal{S}$ and $\mathcal{E}$ together form a valid GPT, then $p_{a|x} {\bf s}_{P^B_{a|x}}\cdot {\bf e}_{b|M^B_y}$ is a probability and we recover Eq.~\eqref{defnB}.

The upper limit on the range of possible values of the CHSH inequality violation among the theories in
$\texttt{GPT}_{\rm candidates}$, which we denote by $\mathcal{B}_{\rm max}$, is defined analogously to $\mathcal{C}_{\rm max}$ in Eq.~\eqref{Cmax}.

Calculating $\mathcal{B}_{\rm max}$ is a difficult optimization problem that involves varying over every pair $(\mathcal{S},\mathcal{E})$ consistent with the experiment, and for each pair implementing the optimization in Eq.~\eqref{BSE}.

Instead of performing this difficult optimization, we will derive an upper bound on $\mathcal{B}_{\rm max}$, denoted $\texttt{UB}(\mathcal{B}_{\rm max})$.  This is achieved in the same manner that the upper bound on $\mathcal{C}_{\rm max}$ was obtained in the previous section, namely, using a qubit-like outer approximation.  

For qubit-like state and effect spaces, it turns out that the maximum violation of the CHSH inequality is the greater of $\frac34$ or the value given for the probability of success in POM in \eqref{BPOMw}. The proof is provided in Appendix~\ref{CHSHqubit}.

Thus, we infer from Eq.~\eqref{numUBpomp2} that 
\begin{equation}
\texttt{UB}(\mathcal{B}_{\rm max}^{}) = 0.8647 \pm 0.0005.\label{numUBCHSH}
\end{equation}

This provides an upper bound on the interval of $\mathcal{B}$ values in which the true value of the maximal CHSH inequality violation lies, as depicted in Fig.~\ref{fg:B}(c).  As noted earlier, our experiment only provides the trivial lower bound $\texttt{LB}(\mathcal{B}_{\rm min})
= \mathcal{B}_{\rm loc}$. Nontrivial
lower bounds have, of course, been provided in previous Bell experiments using photon polarization, such as Ref.~\cite{christensen15}.

\section{Discussion}

We have described a scheme for constraining what GPTs can model a degree of freedom on which one has statistical data from a prepare-and-measure experiment. It proceeds by a tomographic characterization of the GPT states and effects that best represent the preparations and measurements realized in the experiment.  By computing the duals of these, one constrains the possibilities for the true GPT state and effect spaces.  The tomographic scheme is self-consistent in the sense that it does not require any prior characterization of the preparations and measurements.

The rank of the GPT describing the preparations and measurements realized in our experiment can be determined with very high confidence by our method.  
 Because the models we consider have $k(m+n-k)$ parameters, where $k$ is the rank of the model, $m$ is the number of preparations and $n$ is the number of measurements, increasing the rank of the model by 1 
  increases the parameter count by hundreds in the first experiment and by thousands in the second.  
 For this reason, the Akaike information criterion can deliver a decisive verdict against models that have a rank higher than the smallest rank that yields a respectable $\chi^2$ on the grounds that such higher-rank models grossly {\em overfit} the data.

Our experimental results are consistent with the conclusion that in prepare-and-measure experiments, photon polarization acts like a 2-level quantum system, corresponding to a GPT vector space of dimension 4. 

As emphasized in the introduction and Sec.~\ref{sc:tomocomplevidence}, however, any hypothesis concerning the tomographic completeness of a given set of preparations or measurements is necessarily tentative.  Our experiment provided an opportunity for discovering that the cardinality of a tomographically complete set of preparations (measurements) for photon polarization (or equivalently the dimension of the GPT describing them) deviated from our quantum expectations, but it found no evidence of such a dimensional deviation.

Under the assumption that the set of preparations and measurements we realized {\em were} tomographically complete,
the technique we have described provides a means of obtaining experimental bounds on how the shapes of the state and effect spaces might deviate from those stipulated by quantum theory.
We focused in this article on three examples of such deviations, namely, the failure of the no-restriction hypothesis, supra-quantum violations of Bell inequalities, and supra-quantum or sub-quantum violations of noncontextuality inequalities.

Modifications of quantum theory that posit intrinsic decoherence imply unavoidable noise and thereby a failure of the no-restriction hypothesis.  We focused on the volume ratio of $\mathcal{S}_{\rm logical}$ to $\mathcal{S}$ as a generic measure of the failure of the no restriction hypothesis, and we obtained an upper bound on that measure via the volume ratio of $\mathcal{S}_{\rm consistent}$ to $\mathcal{S}_{\rm realized}$.
This provides an upper bound on the degree of noise in any intrinsic decoherence mechanism.

If one makes more explicit assumptions about the decoherence mechanism, one can be a bit more explicit about the bound.  Suppose that the noise that arises from intrinsic decoherence in a prepare-and-measure experiment on photon polarization corresponds to a partially depolarizing map $\mathcal{D}_{1-\epsilon}$ (Eq.~\eqref{eq:depolmap}) where $\epsilon$ is a small parameter describing the strength of the noise, then GPT tomography would find $\mathcal{S}_{\rm realized} \subseteq \mathcal{S}^{v}_{\rm qubit}$ and  
$\mathcal{E}_{\rm realized} \subseteq \mathcal{E}^{v'}_{\rm qubit}$ 
 where $vv' = 1-\epsilon$.  The best qubit-like inner approximations to $\mathcal{S}_{\rm realized}$ and $\mathcal{E}_{\rm realized}$, denoted by $\mathcal{S}^{w_1}_{\rm qubit}$ and $\mathcal{E}^{w_1'}_{\rm qubit}$ in our article, define a lower bound on $vv'$, namely, $w_1 w_1' \le vv'$, and thereby an upper bound on $\epsilon$, namely, $\epsilon \le 1- w_1 w'_1$.  From our second experiment, we obtained the estimate $w_1 w_1' = 0.969 \pm 0.001$, which implies that $\epsilon \le 0.031 \pm 0.001$.

We have also provided experimental bounds on the amount by which the system we studied could yield Bell and noncontextuality inequality violations in excess of their maximum quantum value.

Because violation of each of the inequalities we have considered is related to an advantage for some information-processing task---specifically, parity-oblivious multiplexing and the CHSH game---it follows that our experimental upper bounds on these violations imply an upper bound on the possible advantage for these tasks.  More generally, our techniques can be used to derive limits on advantages for any task that is powered by nonlocality or contextuality.

Our results also exclude deviations from quantum theory that have some theoretical motivation.
For instance, Brassard {\em et al.}~\cite{Brassard06} have shown that communication complexity becomes trivial if one has CHSH inequality violations of $\frac{1}{2}+\frac{1}{\sqrt{6}} \simeq 0.908$ or higher.  If one assumes that this  is the actual threshold at which communication complexity becomes nontrivial (as opposed to being a nonstrict upper bound) and if one endorses the nontriviality of communication complexity as a principle that the true theory of the world ought to satisfy, then one has reason to speculate that the true theory of the world might achieve a CHSH inequality violation somewhere between the quantum bound of 0.8536 and 0.908.   Our experimental bound, however, rules out most of this range of values.

Our experiment also provides a test (and exclusion) of the hypothesis of universal noncontextuality.  In this capacity, it represents a significant improvement over the best previous experiment~\cite{mazurek2016experimental} expecially vis-a-vis 
  what was identified in Ref.~\cite{mazurek2016experimental} to be the greatest weakness of that experiment, namely, the extent of the evidence for the claim that a given set of measurements or preparations should be considered tomographically complete.  Recall that every assessment of operational equivalence among two preparations (measurements)---from which one deduces the nontrivial consequences of universal noncontextuality---rests upon the assumption that one has compared their statistics for a tomographically complete set of measurements (preparations).

The experiment reported in Ref.~\cite{mazurek2016experimental} implemented eight distinct effects and eight distinct states on single-photon polarization and consequently it had the opportunity to discover that a GPT of dimension 4 did not provide a good fit to the data. 
 In other words, the experiment reported in Ref.~\cite{mazurek2016experimental}, just like the experiment reported here, had the opportunity to discover that the cardinality of the tomographically complete sets of effects and states for photon polarization (hence the dimension of the GPT) was not what quantum theory would lead one to expect, via the sort of precision strategy for detecting dimensional deviations described in the introduction and in Section~\ref{sc:tomocomplevidence}.   
  Consequently, it had an opportunity to discover that quantum expectations regarding operational equivalences were also violated.  
  
  The experimental test of noncontextuality reported in the present article, however, improves on that of Ref.~\cite{mazurek2016experimental} insofar as it provided a much better opportunity for detecting dimensional deviations from quantum theory and hence a much better opportunity for uncovering violations of our quantum expectations regarding what sets of preparations and measurements are tomographically complete, the grounds for all assessments of operational equivalences.  In particular, instead of probing just eight states and effects, we probed one hundred of each in the first experiment and one thousand in the second, and then we explicitly explored the possibility that GPT models with rank greater than 4 might provide a better fit to the data.  In particular, we used the Akaike criterion, which incorporates not only the quality of fit of a model ($\chi^2$) but also the number of parameters it requires to achieve this fit, to determine which rank of model is most likely given the data. 

It is important to recall that our experiment probed only a single type of system: the polarization degree of freedom of a photon.  A question that naturally arises at this point is: to what extent can  our conclusions be ported to other types of systems?  

Consider first the question of portability to other types of {\em two-level} systems (by which we mean systems which are described quantumly by a two-dimensional Hilbert space). If it were the case that 
different two-level systems could be governed by different GPTs, this would immediately lead to a thorny problem of how to ensure that the different restrictions on their behaviours were respected even in the presense of interactions between them.  Indeed, the principle that every $n$-level system has the same GPT state and effect spaces as every other has featured in many reconstructions of quantum theory within the GPT framework (see, e.g., the subspace axiom in Ref.~\cite{hardy01}, and its derivation from other axioms in Ref.~\cite{Hardy2016reconstructingQT}) and is taken to be a very natural assumption.  This suggests that there are good theoretical grounds for thinking that our experimental constraints on possible deviations from quantum theory are applicable to {\em all} types of two-level systems. 

It is less clear what conclusions one might draw for $n$-level systems when $n\ne 2$.  
For instance, although quantumly the maximum violation of a CHSH inequality is the same regardless of whether Bob's system is a qubit or a qutrit, this might not be the case for some nonquantum GPT.  Therefore, although there are theoretical reasons for believing that our upper bound on the degree of CHSH inequality violation (assuming no dimensional deviation) applies to all two-level systems, 
we cannot apply those reasons to argue that violations will be bounded in this way for $n$-level systems.  Nonetheless, if one does assume that all two-level systems are described by the same GPT, then we have constraints on the state and effect spaces of every two-level system that is embedded (as a subspace) within the $n$-level system.  This presumably restricts the possibilities for the state and effect spaces of the $n$-level system itself.  How to infer such restrictions---for instance, how to infer an upper bound on the maximal CHSH inequality violation for a three-level system from one on a two-level system---is an interesting problem for future research. 

There is evidently a great deal of scope for further experiments of the type described here.  An obvious direction for future work is to apply our techniques to the characterization of higher dimensional systems and composites.  Another interesting extension would be to generalize the technique to include GPT tomography of transformations, in addition to preparations and measurements.  This is the GPT analogue of quantum process tomography, on which there has been a great deal of work due to its application in benchmarking experimental implementations of gates for quantum computation.  It is likely that many ideas in this sphere can be ported to the GPT context.  A particularly interesting case to consider is the scheme known as {\em gate set tomography}~\cite{merkel13,blumekohout13,greenbaum15}, which achieves a high-precision characterization of a set of quantum gates in a self-consistent manner.

\appendix

\section{Experimental details}

\subsection{Photon source}
The 20 mm long PPKTP crystal is pumped with 0.29 mW of continuous wave laser light at 404.7 nm, producing pairs of 809.4 nm photons with orthogonal polarizations. We detect approximately 22\% of the herald photons produced, and approximately 9\% of the signal photons produced. In order to characterize the single-photon nature of the source we performed a $g^2(0)$ measurement~\cite{grangier86} and found $g^2(0)=0.00184 \pm 0.00003$. This low $g^2(0)$ measurement implies that the ratio of double pairs to single pairs produced by the source is $\sim 1:2000$. We found that if we increased the pump power then a rank 4 model no longer fit the data well. This is because the two-photon state space has a higher dimension than the one-photon state space. The avalanche photodiode single photon detectors we use respond nonlinearly to the number of incoming photons~\cite{resch01}; this makes our measurements sensitive to the multi-pair component of the downconverted light and ultimately limits the maximum power we can set for the pump laser.

\subsection{Measurements}

After a photon exits the measurement PBS, the probability that it will be detected depends on which port of the PBS it exited from. This is because the efficiencies of the two paths from the measurement PBS to the detector are not exactly equal, and also because the detectors themselves do not have the same efficiency. To average out the two different efficiencies we perform each measurement in two stages.  

We will use language from quantum mechanics to explain our procedure. Say we want to perform a projective measurement in the $|\psi\rangle$-$|\psi^\perp\rangle$ basis, for some polarization $|\psi\rangle$ and its orthogonal partner $|\psi^\perp\rangle$. We first rotate our measurement waveplates so they rotate $|\psi\rangle$ to the horizontal polarization, $|H\rangle$ (and thus, $|\psi^\perp\rangle$ is rotated to the vertically polarized state $|V\rangle$). In each output port, we record the number of photons detected in coincidence with the herald, over an integration time of four seconds. We label detections in the transmitted port with `0' and detections in the reflected port with `1'. Second, we rotate the measurement waveplates such that $|\psi\rangle \rightarrow |V\rangle$ and $|\psi^\perp\rangle \rightarrow |H\rangle$. We then swap the labels on the measurement outcomes such that the reflected port corresponds to outcome `0' and the transmitted port to `1'. We again record the number of coincidences between each output port and the herald for four seconds. Finally, we sum the total number of `0' detections, and also the total number of `1' detections over the total eight-second measurement time. The measured frequency at which we obtained outcome `0' is then the total number of `0' detections divided by the sum of the total number of `0' and `1' detections.

\subsubsection{Threefold coincidences}
Sometimes, all three detectors in the experiment fire within a single coincidence window. These events are most likely caused by either a multi-pair emission from the source, or the successful detection of both photons in a single pair in conjunction with a background count at the third detector. We choose to interpret each threefold coincidence as a pair of pairwise coincidences; one between the herald and transmitted port detectors, and one between the herald and reflected port detectors.

Since we are only interested in characterizing the single-pair emissions from our source (and not multi-pair ones), we could have chosen to instead discard all threefold-coincidence events completely. We note that if we had done this, the raw frequency data to which we fit our GPT would change, on average, by an amount that is only 0.01\% of the statistical uncertainty on these frequencies. Using the Akaike information criterion, we would still have concluded that the GPT most likely to describe the data is rank 4. Finally, the probabilities in the rank-4 GPT of best fit would be essentially unchanged, and the shapes of the reconstructed GPT state and effect spaces (and therefore also the inferences made about the achievable inequality violations) would not be affected in any significant way.

\section{Choice of preparation and measurement settings}\label{sc:targetstates}

We choose the preparation and measurement settings in our experiment with the aim of characterizing the largest volume of the state and measurement effect spaces as possible. The state and effect spaces in any GPT are convex, and thus fully characterizing the boundaries of these spaces fully determines the full spaces. Thus our aim is to find preparation and measurement settings that map out the boundaries of the state and effect spaces as best we can, given the finite number of settings we are able to perform.

We use quantum theory to inform our choice of settings. We expect the GPT describing our experiment to be equal to (or very closely approximated by) the GPT for a qubit. The surface of the Bloch sphere (i.e. the space of pure qubit states) determines the qubit state space, and preparing a set states of states that are evenly distributed around the surface of the Bloch sphere should do a good job at characterizing the GPT state space describing our experiment. The qubit effect space is characterized by the surface of the sphere representing projective measurement effects, plus the unit effect, $\mathbb{I}$, and its complement, the zero effect.
Thus, we aim to perform a set of measurements whose effects are evenly distributed on the outside of the sphere of projective effects.

To choose the preparation settings we first find a set of pure quantum states labelled with $|\psi_i\rangle$ that are approximately evenly distributed around the surface of the Bloch sphere. We then find the quarter and half waveplate angles necessary to create each of those states, and each pair of quarter and half waveplate angles is one preparation setting. The space of projective effects is also determined by the Bloch sphere, since every projective effect $|\psi_i\rangle\langle \psi_i|$ can be associated with the state to which it responds deterministically, $|\psi_i\rangle$. The measurement settings are the waveplate angles that implement the projective measurements $\{|\psi_i\rangle\langle \psi_i|, \mathbb{I}-|\psi_i\rangle\langle \psi_i|\}$.

We use a method due to Rakhmanov, Saff, and Zhou~\cite{rakhmanov94} to find the set of approximately uniformly distributed points on the surface of the Bloch sphere. The points lie on a spiral that begins at the south pole of the sphere, and winds up around the sphere and ends at the north pole. The quantum states corresponding to each of the 100 preparation settings in the first experiment are shown in Fig.
10(a),
and the 1000 states corresponding to each preparation setting in the second experiment are displayed in Fig.
10(b).

\begin{figure}
\includegraphics{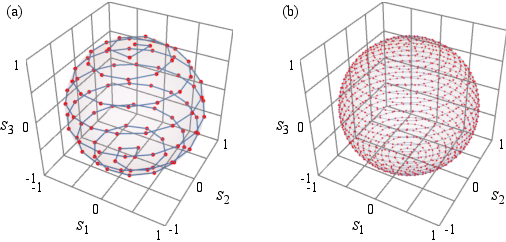}
\caption{(Color). Quantum description of the target states created and measurements performed in our experiment. An evenly distributed set of points lying on a spiral was used to choose the settings for (a) the 100 preparations and measurements characterized in the first experiment and (b) the 1000 nonfiducial preparations and measurements characterized in the second experiment. Each red dot corresponds to a quantum state $|\psi_i\rangle$, and the waveplate angles (i.e., preparation settings) were chosen as those which, under the assumption of the correctness of quantum theory, would prepare those states. Each red dot also defines an effect $|\psi_i\rangle\langle\psi_i|$ which is part of the projective measurement $\{|\psi_i\rangle\langle \psi_i|, \mathbb{I}-|\psi_i\rangle\langle \psi_i|\}$.}
\label{fg:spiral}
\end{figure}

In the second experiment, we also implement a set of six fiducial preparations which we use to characterize each of the 1000 effects in Fig.~10(b),
 and a set of six fiducial measurements which we use to characterize each of the 1000 states in Fig.~10(b).
 The fiducial preparation and measurement sets are shown in Fig.~11.

\begin{figure}
\includegraphics{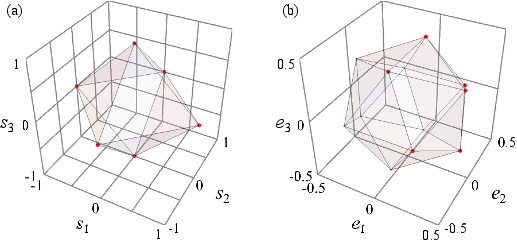}
\label{fg:fiducialprepmeas}
\caption{(Color). Quantum description of the fiducial states and measurement effects performed in the second experiment. (a) Red dots represent the six fiducial states used to characterize the 1000 measurements in Fig.~\ref{fg:spiral}(b). These correspond to the +1 and -1 eigenstates of the three Pauli operators $\sigma_x$, $\sigma_y$, and $\sigma_z$. (b) Red dots represent the six fiducial measurement effects used to characterize each of the states in Fig.~\ref{fg:spiral}(b). These effects lie on six of the twelve vertices of an icosahedron, and they correspond to the outcome-`0' effect of a projective measurement. Each outcome-`0' effect has a corresponding outcome-`1' effect; each outcome-`1' effect is represented by one of the other six vertices on the icosahedron.}
\end{figure}

\section{Finding the rank-$k$ matrix $\tilde{D}$ that best fits the frequency matrix $F$}
\label{ap:fitting}

In this section we explain the algorithm we use to find a low-rank matrix that best fits the matrix of raw frequency data.

For an $m\times n$ matrix of frequency data, $F$, we define the rank-$k$ matrix of best fit, $\tilde{D}$, as the one that minimizes the weighted $\chi^2$ value:
\begin{equation}
	\chi^2 = \sum\limits_{i=1}^m\sum\limits_{j=1}^n\left(\frac{F_{ij} - \tilde{D}_{ij}}{\Delta F_{ij}}\right)^2,
\label{eq:chisq}
\end{equation}
where the weights $\Delta F_{ij}$ are the uncertainties in the measured frequencies, which are calculated assuming Poissonian error in the counts (in cases where we did not collect data for the preparation-measurement pair corresponding to entry $F_{ij}$, we set $\Delta F_{ij} = \infty$). Since $\tilde{D}$ represents an estimate of the true probabilities underlying the noisy frequency data, we need to ensure that $\tilde{D}$ only contains entries between 0 and 1. Hence the matrix of best fit is the one which solves the following minimization problem:
\begin{equation}
 \begin{aligned}\label{eq:chisq_comp}
 & \underset{\tilde{D} \in M_{mn}}{\text{minimize}} & & \chi^2 , \\
 & \text{subject to} & & \rank(\tilde{D}) \leq k  \\
 & & & 0 \leq \tilde{D}_{ij} \leq 1 & \forall \, i,j,
 \end{aligned}
\end{equation}
where $M_{mn}$ is the space of all $m\times n$ real matrices.
The entries in the column of ones (representing the unit measurement effect) that we include in $F$ are {\em exact}, meaning that they have an uncertainty of 0. As $\tilde{D}$ is defined as the matrix that minimizes $\chi^2$, this enforces that the entries in the same column of $\tilde{D}$ will also remain exactly 1. Otherwise, $\chi^2$ would be undefined.

To enforce the rank constraint, we use the parameterization $\tilde{D} = \tilde{S}\tilde{E}$, where $\tilde{S}$ has size $m\times k$ and $\tilde{E}$ is $k \times n$. This minimization problem as stated is NP-hard~\cite{gillis11}, and cannot be solved analytically. However, if either $\tilde{S}$ or $\tilde{E}$ remains fixed, optimizing the other variable is a convex problem which can be solved with quadratic programming. We minimize $\chi^2$ by performing a series of alternating optimizations over $\tilde{S}$ and $\tilde{E}$~\cite{markovsky12}. 

Each iteration begins with an estimate for $\tilde{E}$, and we then consider a variation over the $m \times k$ matrix $\tilde{S}$ such that the $m\times n$ matrix $\tilde{D} = \tilde{S}\tilde{E}$  minimizes the $\chi^2$. Next, we fix $\tilde{S}$  to be the one that achieved the minimum in this variation and we consider a variation over the $k \times n$ matrix $\tilde{E}$ such that $\tilde{D} = \tilde{S}\tilde{E}$ minimizes the $\chi^2$. This is the end of one iteration, and the matrix $\tilde{E}$ that achieved the minimum becomes the $\tilde{E}$ for the beginning of the next iteration. The algorithm runs until a specific convergence threshold is met (i.e., if $\Delta \chi^2 < 10^{-6}$ between successive iterations), or until a maximum number of iterations (we choose 5000) is reached.

We will now show that optimization over $\tilde{S}$ or $\tilde{E}$ is convex (given that the other variable is fixed). For what follows, we will make use of the $\vect(\cdot)$ operator, which takes a matrix and reorganises its entries into a column vector with the same number of entries as the original matrix. For example, given an $m\times n$ matrix $A$, $\vect(A)$ is a vector of length $mn$, and the first $m$ entries of $\vect(A)$ are equal to the first column of $A$, entries $m+1$ through $2m$ are equal to the second column of $A$, and so on. We also define a diagonal $mn\times mn$ matrix of weights, $W$, to encode the uncertainties $(1/\Delta F_{ij})^2$. These values appear along the diagonal of $W$, and they are appropriately ordered such that we can rewrite $\chi^2$ in the more convenient form:
\begin{align}
	\chi^2 	&= \vect(F-\tilde{S}\tilde{E})^T W \vect(F-\tilde{S}\tilde{E}) \\
			&= \vect(\tilde{S}\tilde{E})^T W \vect(\tilde{S}\tilde{E}) - 2 \vect(\tilde{S}\tilde{E})^T W \vect(F) \nonumber \\
			& \quad  + \vect(F)^T W \vect(F),
\label{eq:chisqvec}
\end{align}
where we have also made the substitution $\tilde{D}=\tilde{S}\tilde{E}$.

Defining $I_m$ as the $m\times m$ identity matrix, we can use the identity $\vect(\tilde{S}\tilde{E}) = ( \tilde{E}^T \otimes I_m ) \vect(\tilde{S})$ to write:
\begin{align}
	\chi^2 	&= \vect{(\tilde{S})}^T ( \tilde{E} \otimes I_m ) W ( \tilde{E}^T \otimes I_m ) \vect{(\tilde{S})} \nonumber \\
			& \quad - 2 \vect{(\tilde{S})}^T ( \tilde{E} \otimes I_m ) W \vect(F) \nonumber \\
			& \quad  + \vect(F)^T W \vect(F), 
\label{eq:Poutside}
\end{align}
and we now see that the minimization over $P$ can be written as:
\begin{equation}
 \begin{aligned}\label{eq:Poptim}
 & \underset{\tilde{S} \in M_{mk}}{\text{minimize}} & & \vect{(\tilde{S})}^T ( \tilde{E} \otimes I_m ) W ( \tilde{E}^T \otimes I_m ) \vect{(\tilde{S})} \\
			& & & \quad - 2 \vect{(\tilde{S})}^T ( \tilde{E} \otimes I_m ) W \vect(F) \\
 & \text{subject to} & & 0 \leq (\tilde{S}\tilde{E})_{ij} \leq 1 \,\,\,\,\, \forall \, i,j.
 \end{aligned}
\end{equation}
We have ignored the third term of Eq.~(\ref{eq:chisqvec}) as it is a constant, and depends neither on $\tilde{S}$ nor $\tilde{E}$. Since $W$ is a diagonal matrix consisting of only positive elements, $( \tilde{E} \otimes I_m ) W ( \tilde{E}^T \otimes I_m )$ is positive-definite. This means that (\ref{eq:Poptim}) is a convex quadratic program~\cite{boydvandenberghe2004} which can be solved in polynomial time.

The optimization over $\tilde{E}$ takes a similar form, which can be found by applying the identity $\vect(\tilde{S}\tilde{E}) = ( I_n \otimes \tilde{S} ) \vect(\tilde{E})$ to Eq.~\eqref{eq:chisqvec}:
\begin{equation}
 \begin{aligned}\label{eq:Loptim}
 & \underset{\tilde{E} \in M_{kn}}{\text{minimize}} & & \vect(\tilde{E})^T( I_n \otimes S )^T W ( I_n \otimes \tilde{S} ) \vect(\tilde{E}) \\
			& & & \quad - 2 \vect(\tilde{E})^T( I_n \otimes \tilde{S} )^T W \vect(F) \\
 & \text{subject to} & & 0 \leq (\tilde{S}\tilde{E})_{ij} \leq 1 \,\,\,\,\, \forall \, i,j.
 \end{aligned}
\end{equation}

\section{Decomposition of the fitted matrix of probabilities}
\label{ap:decomp}

As discussed in Section~\ref{sc:estimating} in the main paper, we find a decomposition $\tilde{D}^\mathrm{realized}=\tilde{S}^\mathrm{realized}\tilde{E}^\mathrm{realized}$ in order to characterize the estimates of the spaces realized by the experiment, $\tilde{\mathcal{S}}_{\rm realized}$ and $\tilde{\mathcal{E}}_{\rm realized}$. Here, $\tilde{D}^\mathrm{realized}$ has size $m \times n$, $\tilde{S}^\mathrm{realized}$ is $m \times k$ and $\tilde{E}^\mathrm{realized}$ is $k \times n$. In this appendix we describe the method we use to perform the above decomposition.

We choose the decomposition  to ensure that the first column of $\tilde{S}^\mathrm{realized}$ is a column of ones, which allows us to represent $\tilde{\mathcal{S}}_\mathrm{realized}$ in $k-1$ dimensions. (In our experiment we found $k=4$, but we will use the symbol $k$ in this appendix for generality.) We achieve this by ensuring that the leftmost column in $\tilde{D}^\mathrm{realized}$ is a column of ones representing the unit measurement, such that $\tilde{D}^\mathrm{realized}$ takes the form:
\begin{equation}\label{eq:drealappendix}
\tilde{D}^\mathrm{realized} = \left( \begin{array}{cccccc}
1 & p(0|P_1,M_2) & \cdots  & p(0|P_1,M_n)\\
\vdots & \vdots  & \ddots  & \vdots \\
1 & p(0|P_m,M_2) & \cdots  & p(0|P_m,M_n)\\
\end{array} \right).
\end{equation}

We then proceed to perform the QR decomposition~\cite{lay2002} $\tilde{D}^\mathrm{realized} = QR$, where $R$ is an $m\times n$ upper-right triangular matrix and $Q$ an $m \times m$ unitary matrix. Because $\tilde{D}^\mathrm{realized}$ has the form of Eq.~\eqref{eq:drealappendix}, each entry in the first column of Q will be equal to some constant $c$. We define $Q' = Q/c$ and $R' = cR$, which ensures that the first column of $Q'$ is a column of ones.

Next, we partition $Q'$ and $R'$ as $Q' = \begin{pmatrix} Q_0 & Q_1\end{pmatrix}$ and $R' = \begin{pmatrix} R_0 \\ R_1 \end{pmatrix}$, where $Q_0$ is the first column of $Q'$, $Q_1$ is all remaining columns of $Q'$, $R_0$ is the first row of $R'$, and $R_1$ is all remaining rows of $R'$. We take the singular value decomposition $Q_1 R_1 = U\Sigma V^T$. $Q_1 R_1$ is rank-$(k-1)$, and thus only has $(k-1)$ nonzero singular values. Hence we can partition $U$, $\Sigma$, and $V$ as $U = \begin{pmatrix} U_{k-1} & U_{(k-1)\perp}\end{pmatrix}$, $\Sigma = \begin{pmatrix} \Sigma_{k-1} & 0 \\ 0 & 0\end{pmatrix}$, and $V = \begin{pmatrix} V_{k-1} & V_{(k-1)\perp}\end{pmatrix}$. Here $\Sigma_{k-1}$ is the upper-left $(k-1)\times(k-1)$ corner of $\Sigma$, and $U_{k-1}$ and $V_{k-1}$ are the leftmost $(k-1)$ columns of $U$ and $V$, respectively. Finally, we define $\tilde{S}^\mathrm{realized}$ and $\tilde{E}^\mathrm{realized}$ as $\tilde{S}^\mathrm{realized} = \begin{pmatrix} Q_0 & U_{k-1}\sqrt{\Sigma_{k-1}} \end{pmatrix}$ and $\tilde{E}^\mathrm{realized} = \begin{pmatrix} R_0 \\ \sqrt{\Sigma_{k-1}}V_{k-1}^T \end{pmatrix}$.

The procedure described above ensures that $\tilde{S}^{\rm realized}$ and $\tilde{E}^{\rm realized}$ take the forms:
\begin{equation}
\tilde{S}^{\rm realized} = \left( \begin{array}{cccc}
1 & s_1^{(1)} & \cdots & s_{k-1}^{(1)} \\
1 & s_1^{(2)} & \cdots & s_{k-1}^{(2)} \\
\vdots & \vdots & \ddots & \vdots \\
1 & s_1^{(m)} & \cdots & s_{k-1}^{(m)} \\

\end{array} \right),\label{eq:Sappendix}
\end{equation}
and
\begin{equation}
\tilde{E}^{\rm realized} = \left( \begin{array}{cccc}
1 		& e_0^{(2,0)} 		& \cdots 	& e_0^{(n,0)}\\
0 		& e_1^{(2,0)} 		& \cdots	& e_1^{(n,0)}\\
\vdots 	& \vdots			& \ddots 	& \vdots	 \\
0 		& e_{k-1}^{(2,0)}	& \cdots	& e_{k-1}^{(n,0)}\\
\end{array} \right),
\label{eq:Eappendix}
\end{equation}
where $s_t^{(u)}$ is the $t$-th element of the GPT state vector representing the $u$-th preparation, and 
$e_{t}^{(v,0)}$ is the $t$-th element of the GPT effect vector representing the $0$-th outcome of the $v$-th measurement.

\subsection{Convex closure under convex mixtures and classical post-processing of $\tilde{E}^{\rm realized}$}

As discussed in Section~\ref{sc:estimating}, $\tilde{\mathcal{E}}_{\rm realized}$ is obtained by considering the convex closure under convex mixtures and classical post-processing of $\tilde{E}^{\rm realized}$. We only perform two-outcome measurements in our experiment, and thus the full set of effects in $\tilde{\mathcal{E}}_{\rm realized}$ is the convex hull of the outcome-0 effects of all measurement procedures implemented in the experiment (i.e. the matrix $\tilde{E}^{\rm realized}$) {\em and} of all the outcome-1 effects of all the implemented measurements (i.e the matrix 1-$\tilde{E}^{\rm realized}$).

If we chose to, we could simply take the $\tilde{E}^{\rm realized}$ returned by the decomposition of $\tilde{D}^{\rm realized}$ that we described above, and define the larger matrix $\begin{pmatrix} \tilde{E}^{\rm realized} & 1-\tilde{E}^{\rm realized}\end{pmatrix}$, and the convex hull of the vectors in this larger matrix would define our estimate, $\tilde{\mathcal{E}}_{\rm realized}$, of the space of GPT effects realized in the experiment.

However, in an attempt to treat the outcome-0 and outcome-1 effect vectors on equal footing, we instead define the larger matrix $\tilde{D}^{\rm R} = \begin{pmatrix} \tilde{D}^{\rm realized} & 1-\tilde{D}^{\rm realized}\end{pmatrix}$. We then find a decomposition $\tilde{D}^{\rm R} = \tilde{S}^{\rm realized}\tilde{E}^{\rm R}$ using the method described above. This ensures that $\tilde{E}^{\rm R}$ has the form:
\begin{equation}
\tilde{E}^{\rm R} = \left( \begin{array}{cccccccc}
1 		& e_0^{(2,0)} 		& \cdots 	& e_0^{(n,0)} 	& 0			& e_0^{(2,1)}	& \cdots	& e_0^{(n,1)}\\
0 		& e_1^{(2,0)} 		& \cdots	& e_1^{(n,0)} 	& 0 		& e_1^{(2,1)} 	& \cdots	& e_1^{(n,1)}				\\
\vdots 	& \vdots			& \ddots 	& \vdots		& \vdots	& \vdots 		& \ddots	& \cdots  \\
0 		& e_{k-1}^{(2,0)}	& \cdots	& e_{k-1}^{(n,0)}	& 0			& e_{k-1}^{(2,1)}	& \cdots	& e_{k-1}^{(n,1)}\\

\end{array} \right).\label{eq:Eappendix2}
\end{equation}

\section{Calculation of dual spaces}\label{sc:dualcalc}

The spaces $\mathcal{\tilde{S}}_{\rm consistent}$ and $\mathcal{\tilde{E}}_{\rm consistent}$ are the duals of the realized spaces $\mathcal{\tilde{E}}_{\rm realized}$ and $\mathcal{\tilde{S}}_{\rm realized}$, respectively. Here we will discuss how we calculate the consistent spaces from the realized ones.

We start with the calculation of $\mathcal{\tilde{S}}_{\rm consistent}$. By definition, $\mathcal{\tilde{S}}_{\rm consistent}$ is the intersection of the geometric dual of $\mathcal{\tilde{E}}$ and the set of all normalized GPT states; specifically, the set of $\mathbf{s} \in \mathbb{R}^k$ such that $\forall \mathbf{e}  \in \mathcal{\tilde{E}}_{\rm realized}: 0 \leq \mathbf{s} \cdot  \mathbf{e} \leq 1$ and such that  $\mathbf{s}\cdot\mathbf{u}=1$. This definition (called an {\em inequality representation}) completely specifies $\mathcal{\tilde{S}}_{\rm consistent}$. However, in order to perform transformations on the space or calculate its volume, it can be useful to have its {\em vertex description} as well, which is a list of vertices that completely specify the space's convex hull. Finding a convex polytope's vertex representation given its inequality representation is called the {\em vertex enumeration problem}~\cite{avis92}.

To find the vertex representation of $\mathcal{\tilde{S}}_{\rm consistent}$, we first simplify its inequality representation. Since $\mathcal{\tilde{E}}_{\rm realized}$ is a convex polytope, we don't need to consider every $\mathbf{e}$ in $\mathcal{\tilde{E}}_{\rm realized}$, but only the vertices of $\mathcal{\tilde{E}}_{\rm realized}$.
If we denote the set of vertices of $\mathcal{\tilde{E}}_{\rm realized}$ by ${\rm Vertices}\left(\mathcal{\tilde{E}}_{\rm realized}\right)$,
 then we can replace the $\forall \mathbf{e}  \in \mathcal{\tilde{E}}_{\rm realized}$ in the definition of $\mathcal{\tilde{S}}_{\rm consistent}$ with $\forall \mathbf{e}  \in {\rm Vertices}\left(\mathcal{\tilde{E}}_{\rm realized}\right)$. Calculation of ${\rm Vertices}\left(\mathcal{\tilde{E}}_{\rm realized}\right)$ is performed with the pyparma~\cite{pyparma} package in Python 2.7.6. The calculation of the vertex description of $\mathcal{\tilde{S}}_{\rm consistent}$ is performed with an algorithm provided by Avis and Fukuda~\cite{avis92}. We use functions in pyparma~\cite{pyparma} which call the cdd library~\cite{cddlib} to find the vertex description of $\mathcal{\tilde{S}}_{\rm consistent}$.

Finding the vertex description of $\mathcal{\tilde{E}}_{\rm consistent}$ from $\mathcal{\tilde{S}}_{\rm realized}$ is done in an analogous way. $\mathcal{\tilde{E}}_{\rm consistent}$ is defined as the geometric dual of the space that is the subnormalization of $\tilde{\mathcal{S}}_{\rm realized}$, $\{ w\mathbf{s}:\mathbf{s}\in\tilde{\mathcal{S}}_{\rm realized}, w\in [0,1]\}$. The subnormalization of $\tilde{\mathcal{S}}_{\rm realized}$ is also the convex hull of the union of the GPT state vectors that make up the rows of $\tilde{S}^{\rm realized}$ and the GPT state vector with $s_0=\dots=s_{k-1}=0$ that represents the state with normalization zero.

\section{Maximal CHSH inequality violations with qubit-like state spaces}\label{CHSHqubit}

We here provide a proof of the fact that the optimal value of the CHSH inequality when Bob's system is described by a qubit-like state and effect space is the same as the value of the POM noncontextuality inequality for the same case, 
provided that the latter is at least $\frac34$, that is,
\begin{equation}\label{F1}
\mathcal{B}_{(\mathcal{S}^{w}_{\rm qubit},\mathcal{E}^{w'}_{\rm qubit})}^{} 
= \max\left\{\frac34 ,\mathcal{C}_{(\mathcal{S}^{w}_{\rm qubit},\mathcal{E}^{w'}_{\rm qubit})}\right\}.
\end{equation}

We begin with a geometric charachterization of $\mathcal{S}^w_{\rm qubit}$ and $\mathcal{E}^{w'}_{\rm qubit}$. Recalling the Bloch representation of $\mathcal{S}_{\rm qubit}$ and $\mathcal{E}_{\rm qubit}$ from Sec.~\ref{examples}, and noting that the maximally mixed state is represented by $(1,0,0,0)$, applying $\mathcal{D}_w$ from Eq.~\eqref{eq:depolmap} gives $\mathcal{S}^{w}_{\rm qubit}$ as a ball of radius $w$, i.e. $(1,s_1, s_2, s_3)$ with $\sqrt{s_1^2+s_2^2+s_3^2} \leq w$. Similarly $\mathcal{E}^{w'}_{\rm qubit}$ is a ``Bloch diamond'' with radius $w'$, i.e., $(e_0, e_1, e_2, e_3)$ with $0 \leq e_0 \leq 1$ and $\sqrt{e_1^2+e_2^2+e_3^2} \leq w'\min\{e_0, 1-e_0\}$.

In particular, $\mathcal{E}^{w'}_{\rm qubit}$ is the convex hull of $(0,0,0,0)$, $(1,0,0,0)$ and effects of the form $\left(\frac12, e_1, e_2, e_3\right)$ with $\sqrt{e_1^2+e_2^2+e_3^2} = \frac12 w'$. Thus this GPT shares with a qubit the feature that all binary-outcome measurements are convex combinations of (the analog of) projective measurements.  Specifically, the extremal binary-outcome measurements consist of the {\em trivial} binary-outcome measurement with effects $(0,0,0,0)$ and $(1,0,0,0)$, and the {\em nontrivial} binary-outcome measurements with effects $\left(\frac12, e_1, e_2, e_3\right)$ and $\left(\frac12, -e_1,- e_2, -e_3\right)$  with $\sqrt{e_1^2+e_2^2+e_3^2} = \frac12 w'$.

Recall from Eq.~\eqref{BSE} that we are interesting in maximizing 
\begin{equation}
\frac14 \sum_{a,b,x,y} \delta_{a\oplus b, xy} p_{a|x} {\bf s}_{P^B_{a|x}}\cdot {\bf e}_{b|M^B_y},
\label{BSE2}
\end{equation}
over $\{p_{a|x}\}, \{{\bf s}_{P^B_{a|x}}\}$ that satisfy the no-signalling constraint, Eq.~\eqref{eq:nosignal}, and over $\{{\bf e}_{b|M^B_y}\}$.  

For each $b$, Eq.~\eqref{BSE2} is convex-linear in Bob's effects ${\bf e}_{b|M^B_y}$. Hence it suffice to maximize Eq.~\eqref{BSE2} over the convexly extremal binary-outcome measurements.
In particular, Bob's optimal strategy will be one of two possibilities: at least one of his measurements is trivial, or both of his measurements are nontrivial.

First, consider the case where the optimum is achieved when one of Bob's measurements is trivial, i.e., has effects $(0,0,0,0)$ and $(1,0,0,0)$. Clearly this measurement can be implemented jointly with any other measurement, regardless of whether this other measurement is trivial or not.  But violating a bipartite Bell inequality such as CHSH requires that both parties use incompatible measurements \cite{fine1982b}. Hence the maximum value of Eq.~\eqref{BSE2} for this case cannot exceed $\mathcal{B}_{\rm loc} = \frac34$. Indeed this value can be achieved with both of Bob's measurements being trivial, for example by having Alice and Bob always output $a = b = 0$.  Therefore, in this case 
\begin{equation}\label{F3}
  \mathcal{B}_{(\mathcal{S}^{w}_{\rm qubit},\mathcal{E}^{w'}_{\rm qubit})} = \frac34.
\end{equation}

Now consider the case where the optimum is achieved when both of Bob's measurements are nontrivial, i.e., for each $(b, y)$, ${\bf e}_{b|M^B_y} = \left(\frac12, e_1, e_2, e_3\right)$ with $\sqrt{e_1^2+e_2^2+e_3^2} = \frac12 w'$.  If we define $\tilde{\bf e}_{b|M^B_y} := \frac{1}{w'}(e_1, e_2, e_3)$, then $\tilde{\bf e}_{b|M^B_y} $ is a vector of length $\frac12$, which---according to the convention we are using in this article (see footnote \ref{blochconvention})---is what one has quantumly. Similarly, because for each $(a,x)$, ${\bf s}_{P^B_{a|x}} = (1, s_1, s_2, s_3)$ with $\sqrt{s_1^2+s_2^2+s_3^2} \leq w$, if we define $\tilde{\bf s}_{P^B_{a|x}} := \frac{1}{w}(s_1,s_2,s_3)$, then $\tilde{\bf s}_{P^B_{a|x}}$ has length at most $1$, which is what one has quantumly. Noting that $\sum_{a,b,x,y} \delta_{a \oplus b,xy}p_{a|x} = \sum_{a,x,y}p_{a|x} = \sum_{x,y}1 = 4$, we have that Eq.~\eqref{BSE2} becomes
\begin{equation}
\frac12 + w w'\frac14 \sum_{a,b,x,y} \delta_{a\oplus b, xy} p_{a|x} \tilde{\bf s}_{P^B_{a|x}}\cdot \tilde{\bf e}_{b|M^B_y}.
\label{BSE3}
\end{equation}
Furthermore,  the no-signalling constraint Eq.~\eqref{eq:nosignal} can be written as
\begin{equation}\label{eq:nosignal2}
  p_{0|0} \tilde{\mathbf{s}}_{P^B_{0|0}} + p_{1|0} \tilde{\mathbf{s}}_{P^B_{1|0}}   
  = p_{0|1} \tilde{\mathbf{s}}_{P^B_{0|1}} + p_{1|1} \tilde{\mathbf{s}}_{P^B_{1|1}}.
\end{equation}
In the case $ww' = 1$, we recover the usual problem of maximizing the CHSH value where Bob does projective measurements on a qubit, for which the maximum value $\mathcal{B}_{\rm Q}$ is given in Eq.~\eqref{eq:tsirelson}. 
(The fact that we can optimize over the ensembles of states to which Alice steers rather than optimizing over the bipartite state and Alice's measurements follows from
the Schr\"{o}dinger-HJW theorem \cite{schr,HJW}.) Since the only place that $w$ and $w'$ appear in the problem is before the sum in Eq.~\eqref{BSE3}, and since $ww' > 0$, it is clear that an optimal strategy for our problem will use the same $p_{a|x}$, $\tilde{\bf s}_{P^B_{a|x}}$ and $\tilde{\bf e}_{b|M^B_y}$ as in the $ww'=1$ case.  Hence, if the optimal strategy uses
 a pair of nontrivial measurements, then
\begin{equation}
\left( \mathcal{B}_{(\mathcal{S}^{w}_{\rm qubit},\mathcal{E}^{w'}_{\rm qubit})} 
  - \frac12 \right) = ww'\left( \mathcal{B}_{\rm Q} - \frac12 \right),
\end{equation}
giving
\begin{eqnarray}
  \mathcal{B}_{(\mathcal{S}^{w}_{\rm qubit},\mathcal{E}^{w'}_{\rm qubit})} &= \frac12 + ww'\frac{1}{2\sqrt2} \\
  &=   \mathcal{C}_{(\mathcal{S}^{w}_{\rm qubit},\mathcal{E}^{w'}_{\rm qubit})},\label{BisC}
\end{eqnarray}
where we have used Eq.~\eqref{BPOMw}.  

It follows that the optimal strategy achieves the maximum of Eq.~\eqref{F3} and Eq.~\eqref{BisC}, which establishes Eq.~\eqref{F1}.

\begin{acknowledgments}
The authors thank Jean-Philippe MacLean, Patrick Daley and Bel\'{e}n Sainz for fruitful discussions. We also thank Jean-Philippe MacLean for assistance with figures, David Schmid for helpful comments on the manuscript. This research was supported in part by the Natural Sciences and Engineering Research Council of Canada (NSERC), Canada Research Chairs, Ontario Centres of Excellence, Industry Canada, the Canada Foundation for Innovation (CFI), and the Royal Commission for the
Exhibition of 1851. Research at Perimeter Institute is supported by the Government of Canada through Industry Canada and by the Province of Ontario through the Ministry of Research and Innovation.
\end{acknowledgments}

\end{document}